\documentclass[aps,prl,twocolumn,superscriptaddress,longbibliography]{revtex4}
\usepackage{graphicx}
\usepackage{dcolumn}
\usepackage{bm}
\usepackage{physics}
\usepackage{amsmath}
\usepackage{amssymb}
\usepackage{array}
\usepackage{color}
\usepackage{times}
\newcolumntype{P}[1]{>{\centering\arraybackslash}p{#1}}
\usepackage[colorlinks=true, breaklinks=true, linkcolor=blue, citecolor=blue, urlcolor=blue]{hyperref}

\newcommand{\bonnpi}{Physikalisches Institut, University of Bonn, Nussallee 12, 53115 Bonn, Germany}
\newcommand{\geneva}{Department of Quantum Matter Physics, University of Geneva, Quai Ernest-Ansermet 24, 1211 Geneva, Switzerland}
\newcommand{\saarbrucken}{Theoretische Physik, Universit\"at des Saarlandes, Campus E26, D-66123 Saarbr\"ucken, Germany}

\begin{document}

\title{Controlling the dynamics of atomic correlations via the coupling to a dissipative cavity}
\date{\today}

\begin{abstract} 
We analyze the relaxation dynamics in an open system, composed by a quantum gas of bosons in a lattice interacting via both contact and global interactions.
We report the onset of periodic oscillations of the atomic coherences exhibiting hallmarks of synchronization after a quantum quench.
The dynamical behavior exhibits the many-body collapse and revival of atomic coherences and emerges from the interplay of the quantum dissipative nature of the cavity field and the presence of a (approximate) strong symmetry in the dissipative system.
We further show that the approximate symmetry can dynamically self-organize. 
We argue that the approximate symmetry can be tailored to obtain long-lived coherences.
These insights provide a general recipe to engineer the dynamics of globally-interacting systems.
\end{abstract}

\author{Catalin-Mihai Halati}
\affiliation{\geneva}
\author{Ameneh Sheikhan}
\affiliation{\bonnpi}
\author{Giovanna Morigi}
\affiliation{\saarbrucken}
\author{Corinna Kollath}
\affiliation{\bonnpi}

\maketitle

Open system and measurement control have attracted an enormous interest in the last decade for the engineering of many-body quantum systems \cite{DiehlZoller2008, VerstraeteCirac2009,MuellerZoller2012, Wiseman2009, ZhangNori2017,PuenteRizzi2024}.
Most proposals target the creation of interesting steady states, e.g. topological states of fermionic matter \cite{BardynDiehl2013}, non-trivial transport properties \cite{LandiSchaller2022}, quantum phases stemming from long-range spin interactions \cite{MivehvarPiazza2019, OstermannMivehvar2019,MasalaevaRitsch2021, ChiocchettaDiehl2021, UhrichHauke2023}, or exhibiting dynamical synthetic gauge fields \cite{KollathBrennecke2016, SheikhanKollath2016, BallantineKeeling2017, HalatiKollath2017, MivehvarPiazza2017, ColellaRitsch2019}.
Much less attention has been devoted to the design of environments affecting the dynamical properties of a quantum system \cite{PolettiKollath2013, SciollaKollath2015, MacieszczakGarrahan2016, delCampoKim2019, KingMorigi2024}. 
In this case, the dynamics of correlations needs to be carefully considered \cite{SciollaKollath2015}.
For example, even though the BCS superconducting state itself can be prepared using dissipative dynamics \cite{YiZoller2012}, nevertheless, the desired superconducting current-current correlations are not present as long as the dissipative coupling is applied \cite{SciollaKollath2015}.

\begin{figure}[!hbtp]
\centering
\includegraphics[width=0.48\textwidth]{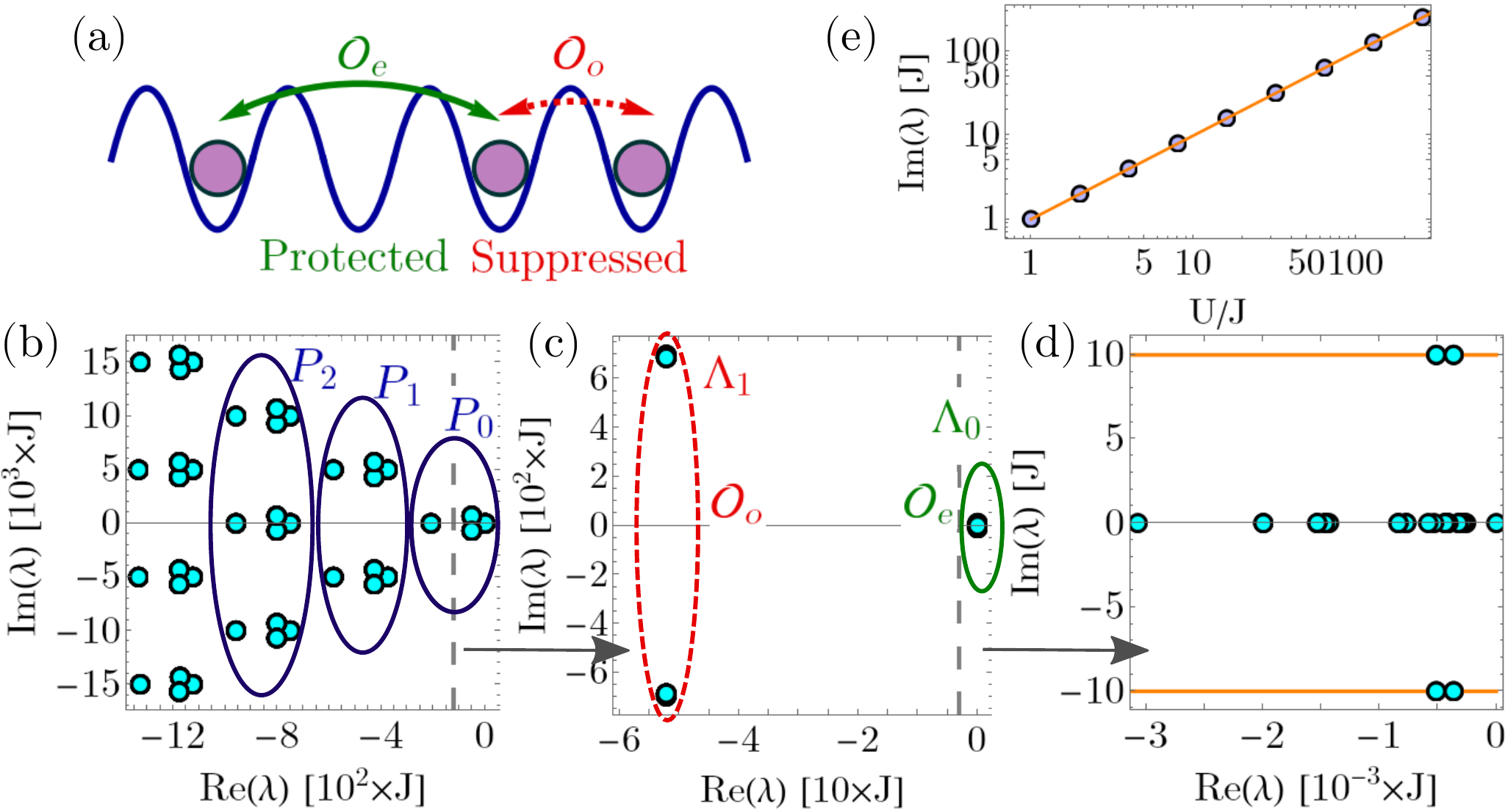}
\caption{
(a) Sketch of ultracold atoms in an optical lattice potential. The operator $\mathcal{O}_e$ probes the coherence between sites at even distance and $\mathcal{O}_o$ at odd distance.
(b)-(d) Eigenvalues spectra of the Liouvillian modeling the Bose-Hubbard model coupled to a dissipative cavity mode, Eq.~(\ref{eq:Lindblad}),  obtained with ED for $L=4$ sites, $N=2$ particles, $\hbar\Omega\sqrt{N}/J=1323$, $\hbar\delta/J=5000$, $\hbar\Gamma/J=750$, $U/J=10$. 
We show the lowest (b) 1000 (c) 50 (d) 34 eigenvalues, where panels (c) and (d) are zoom ins of (b) and (c) at the right of the vertical dashed gray lines (as depicted by the gray arrows).
$P_n$ marks the subspaces containing $n$ photonic excitations, with $\Lambda_0$ and $\Lambda_1$ corresponding to the decoherence free subspace and the first excited subspace for vanishing $J$.
$\mathcal{O}_e$ couples mainly to states in $\Lambda_0$ and $\mathcal{O}_o$ to states in $\Lambda_1$.
(e) The dependence on the on-site interaction $U$ of the imaginary part of the lowest eigenvalue whose imaginary part is in the range $[0.75U,1.25U]$.
}
\label{fig:sketch_ED}
\end{figure}

In this work, we present a general recipe on how to use the intricate interplay of dissipation and symmetries to engineer intriguing metastable and dynamical phenomena in open quantum systems. 
We exemplify this by designing long-lived synchronized oscillations of interacting bosonic atoms coupled to an optical cavity.
The realization of long-lived coherences relies on employing dissipative state engineering and protecting the dynamics via strong symmetries, being related to purely imaginary eigenvalues of the Liouvillian operator, i.e.~rotating coherences \cite{AlbertJiang2014}.  
The coupling between atoms and cavity selectively stabilizes the atomic correlations, which can exhibit synchronization \cite{BucaJaksch2019b,BucaJaksch2022}. 
In the considered example, we select a spatio-temporal pattern such that
the coherences between sites at even distances exhibit long-lived oscillations, while the coherences at odd distances are strongly suppressed [see sketch, Fig.~\ref{fig:sketch_ED}(a)]. 
We show that the quantum nature of the cavity field is essential in determining this dynamics and that the self-organization of the approximate symmetry can lead to a similar behavior.

One important element for understanding and tailoring the dissipative dynamics is the spectrum of the Liouvillian governing the evolution of the density matrix.
The dissipative processes determine the complex nature of the eigenvalues, which in the case of large dissipation rates $\Gamma$, are clustered in bands, with gaps between the real parts proportional to $\Gamma$. 
However, in many-body and hybrid systems, the situation can be much more complex and eigenstates can exist with decay rates smaller than $\Gamma$. 
For an atomic gas coupled to a lossy bosonic mode the Liouvillian $\mathcal{L}$ is given by \cite{Carmichaelbook,BreuerPetruccione2002}
\begin{align}
\label{eq:Lindblad}
& \pdv{t} \rho=\mathcal{L}\rho= -\frac{i}{\hbar} \left[H, \rho \right] + \frac{\Gamma}{2}\left(2a\rho a^\dagger-a^\dagger a \rho-\rho a^\dagger a\right),
\end{align}
where $a$ is the annihilation operator for the bosonic mode.
An exemplary spectrum of $\mathcal{L}$, when $H$ describes a Bose-Hubbard model coupled to a dissipative cavity, is shown in Fig.~\ref{fig:sketch_ED}(b), determined using exact diagonalization (ED).
Due to the direct dissipative coupling to the photon losses, most eigenstates of $\mathcal{L}$ have eigenvalues whose real part is $\propto\Gamma$, e.g.~in the subspaces $P_{1}$ and $P_{2}$, signaling an exponential decay of their contribution to the time dependent density matrix.
Instead, subspace $P_0$ contains coherent photonic states that do not couple directly to dissipation and have a metastable nature, protected from the fast decay. 
Dissipative engineering \cite{MuellerZoller2012} often employs the decoherence free subspace $\Lambda_0$ (subspace of $P_0$), i.e.~corresponding to eigenvalues with vanishing real parts.

Generically, the Hamiltonian in Eq.~(\ref{eq:Lindblad}) can be written as $H=H_\text{c}+H_\text{ac}+H_\text{a}^{(1)}+H_\text{a}^{(2)}$, where $H_\text{c}$ describes the free bosonic mode, $H_\text{ac}\propto(a+a^\dagger)\Delta$ contains the coupling of the mode to an atomic operator $\Delta$, the other terms describing the atomic processes include $H_\text{a}^{(1)}$, which commutes with $\Delta$, and $H_\text{a}^{(2)}$, which does not commute with $\Delta$.
In the limit of vanishing $H_\text{a}^{(2)}$, exciting possibilities arise to engineer dynamical features within the metastable subspace $P_0$, as in its absence a strong symmetry is present \cite{AlbertJiang2014, BucaProsen2012}.
The symmetry generator $\Delta$ commutes with both the Hamiltonian and the jump operator.
This gives another handle on controlling the dynamics, as the evolution in distinct symmetry sectors is independent, each with its own steady or rotating states.
Within $P_0$, the subspace $\Lambda_1$ is generated by coherences between eigenstates of $\Delta$ at different eigenvalues \cite{supp}, while $\Lambda_0$ contains steady states and coherences between degenerate eigenstates of $\Delta$.
Furthermore, $H_\text{a}^{(1)}$ can lift the degeneracy within the eigenspaces of $\Delta$. As a result, the Liouvillian can exhibit purely imaginary eigenvalues \cite{Albert2019, supp}, offering a route for realizing synchronization.
Thus, by designing the dissipative coupling, here via $H_\text{ac}$, we can choose which dynamical features are rapidly suppressed [e.g.~in Fig.~\ref{fig:sketch_ED}(c) correlations probing the excited subspace $\Lambda_1$] and which are protected up to long times (with dynamics dominated by the lowest subspace $\Lambda_0$).
For example, the expectation value of an operator $\mathcal{O}_o$ coupling only distinct eigenstates of $\Delta$ [subspace $\Lambda_1$ in Fig.~\ref{fig:sketch_ED}(c)] experiences fast oscillations and rapid decay to its steady value, compared to $\mathcal{O}_e$, which only couples degenerate eigenstates of $\Delta$ [subspace $\Lambda_0$ in Fig.~\ref{fig:sketch_ED}(c)-(d)].

In the presence of the strong symmetry to probe the dynamics of the different symmetry sector one needs to carefully prepare the initial state. 
To circumvent this problem, we consider a finite contribution from $H_\text{a}^{(2)}$, which breaks the symmetry.
If one slightly breaks a strong symmetry \cite{HalatiKollath2022}, it generally reduces the number of steady states, giving rise to slowly decaying states forming the subspace $\Lambda_0$ within $P_0$.
The decay timescale of these symmetry protected metastable states depends on the magnitude of the symmetry breaking term and can potentially be much smaller than the dissipative gaps of the Liouvillian in the presence of the strong symmetry. 
For example, the states shown in Fig.~\ref{fig:sketch_ED}(e) would have zero real part in the limit $H_\text{a}^{(2)}=0$, however, even with a finite contribution their real parts are still much smaller than the gap to $\Lambda_1$.
We envision these approximate strong symmetries as a tool for the design of metastable states of the Liouvillian \cite{Garcia-RipollCirac2009,ZanardiCamposVenuti2014, MacieszczakGarrahan2016, JinMa2024}, where we can control their lifetime by the magnitude of the symmetry breaking term.
Thus, we consider a protocol in which we start from the ground state of $H_\text{a}^{(1)}+H_\text{a}^{(2)}$, containing the coherences of interest and quench the coupling to the bosonic mode, considering a separation of scales with respect to the atomic processes. 
This induces the dissipative dynamics in the presence of the approximate strong symmetry.

We exemplify the recipe for a one-dimensional lattice of interacting bosonic atoms inside a high finesse cavity, transversely pumped with a standing-wave laser beam and exhibiting photon losses \cite{MaschlerRitsch2008,RitschEsslinger2013,MivehvarRitsch2021} . This showcases a dissipative many-body realization of the collapse and revival of coherences. Such dynamics has been discussed in closed systems for both matter and light fields \cite{Cummings1965, EberlySanchez-Mondragon1980,WrightGarrison1996, DooleySpiller2013}, and has been observed for the matter field of a Bose-Einstein condensate \cite{GreinerBloch2002b}, or nuclear spins \cite{GoldblattWood2024}.
For our example, the Hamiltonian contains $H_{\text{c}}= \hbar\delta a^\dagger a$, with $\delta$ the cavity-pump detuning.
The period of the lattice is chosen to be twice the period of the cavity mode, such that the cavity effectively couples to the even-odd  atomic density imbalance, $H_{\text{ac}}=  -\hbar\Omega ( a + a^\dagger)\Delta,~\Delta=\sum_{j} (-1)^j n_j$, with the coupling strength $\Omega$. 
The coupling commutes with the atomic repulsive on-site interactions, $H_\text{a}^{(1)}\equiv H_{\text{int}}=\frac{U}{2} \sum_{j=1}^L n_{j}(n_{j}-1)$, of strength $U$, and competes with the kinetic processes, $H_\text{a}^{(2)}\equiv H_{\text{kin}}=-J \sum_{j=1}^{L-1} (b_{j}^\dagger b_{j+1} + \text{H.c.})$, with amplitude $J$.
Interacting bosonic lattice models coupled to an optical cavity have been realized experimentally \cite{KlinderHemmerich2015b, LandigEsslinger2016, HrubyEsslinger2018}, while theoretical studies focused mostly on steady state properties  \cite{RitschEsslinger2013,MivehvarRitsch2021,  NiedenzuRitsch2010, SilverSimons2010, VidalMorigi2010,LiHofstetter2013, BakhtiariThorwart2015,FlottatBatrouni2017, ChiacchioNunnenkamp2018,LinLode2019, HimbertMorigi2019, HalatiKollath2020, BezvershenkoRosch2021, HalatiKollath2022, SharmaMorigi2022, ChandaMorigi2022}.

We analyze the quench scenario with the atoms in their ground state and the atoms-cavity coupling suddenly turned on.
We perform the exact time evolution of Eq.~(\ref{eq:Lindblad}) using a recently developed method based on time-dependent matrix product states (tMPS), see Refs.~\cite{supp, HalatiKollath2020b}. 
We complement our understanding with analytical results in the limit $J\to 0$ \cite{supp, HalatiKollath2020b} and ED for small systems. 
These approaches go beyond the often employed mean-field treatment of the cavity-atoms coupling \cite{MivehvarRitsch2021}.
To emphasize the role of the cavity field, we contrast our tMPS results of the atom-cavity system with the dynamics of a Bose-Hubbard model in the presence of a classical light field, i.e.~a superlattice potential.
The superlattice potential $V(t)$ can be obtained as a mean-field description of the coupled dynamics, Eq.~(\ref{eq:Lindblad}), when the cavity is assumed to be a coherent state. 
The Hamiltonian in this situation is given by $H_\text{MF}=H_{\text{int}}+H_{\text{kin}}-V(t)\Delta$ \cite{supp}, and we refer to it as the classical field approach.

\emph{Results in the presence of the approximate strong symmetry:}
Our analysis begins in the regime of vanishing tunneling $J=0$, where analytical results can be obtained \cite{supp}. 
This is motivated by the strong symmetry arising at $J=0$, as local densities are conserved quantities, commuting with both $H$ and $a$ \cite{AlbertJiang2014, BucaProsen2012}. 
These results provide crucial information to our understanding also at small finite $J$, where the sectors of the symmetry are a good approximate description.
We show the ED spectra of $\mathcal{L}$ in Fig.~\ref{fig:sketch_ED}(b)-(d) for a small system, for parameters similar to the experiment performed in \cite{LandigEsslinger2016}.

The subspaces $P_n$, Fig.~\ref{fig:sketch_ED}(b), correspond to excitations on top of the photonic coherent state, for which the main contribution to the real part is given by $n\hbar\Gamma/2$ and by $\hbar\delta$ to the imaginary part. 
The subspaces with $n>0$ show a fast decay and are important only for the short time dynamics. 
Therefore, we focus the analysis on $P_0$, and in particular on the lowest two subspaces $\Lambda_1$ and $\Lambda_0$, Fig.~\ref{fig:sketch_ED}(c). 
Here the photons are in a coherent state determined by the atomic density distribution.
$\Lambda_1$ contains excited states capturing the coherence between different atomic distribution characterized by imbalances $\Delta$ and $\Delta\pm 2$. 
These coherences decay with a rate depending, at large dissipation strength, inversely on $\Gamma$ \cite{supp}, as known from the Zeno effect.
In contrast, $\Lambda_0$ consists of eigenstates with vanishing real part, protected by the strong symmetry for $J=0$. 
As detailed in \cite{supp}, there are several types of states in $\Lambda_0$, steady states of the form $\rho_{0,st}=\ket{\alpha(\Delta);\{n_j\}}\bra{\alpha(\Delta);\{n_j\}}$, or traceless coherences $\rho_0=\ket{\alpha(\Delta);\{n_j\}}\bra{\alpha(\Delta);\{n'_j\}}$ between states with different density distribution and the same odd-even imbalance. 
When the latter describes a coherence between states with different interaction energies, its eigenvalue has a finite imaginary part [orange line in Fig.~\ref{fig:sketch_ED}(d)].
Such states are called rotating coherences and lead to persistent synchronized oscillations in the long-time limit \cite{AlbertJiang2014, BucaJaksch2019b,BucaJaksch2022}.
We checked the dependence of the imaginary part on $U$ for the ED results for small $J$ in Fig.~\ref{fig:sketch_ED}(e), recovering the linear dependence expected for $J=0$.

\begin{figure}[!hbtp]
\centering
\includegraphics[width=0.48\textwidth]{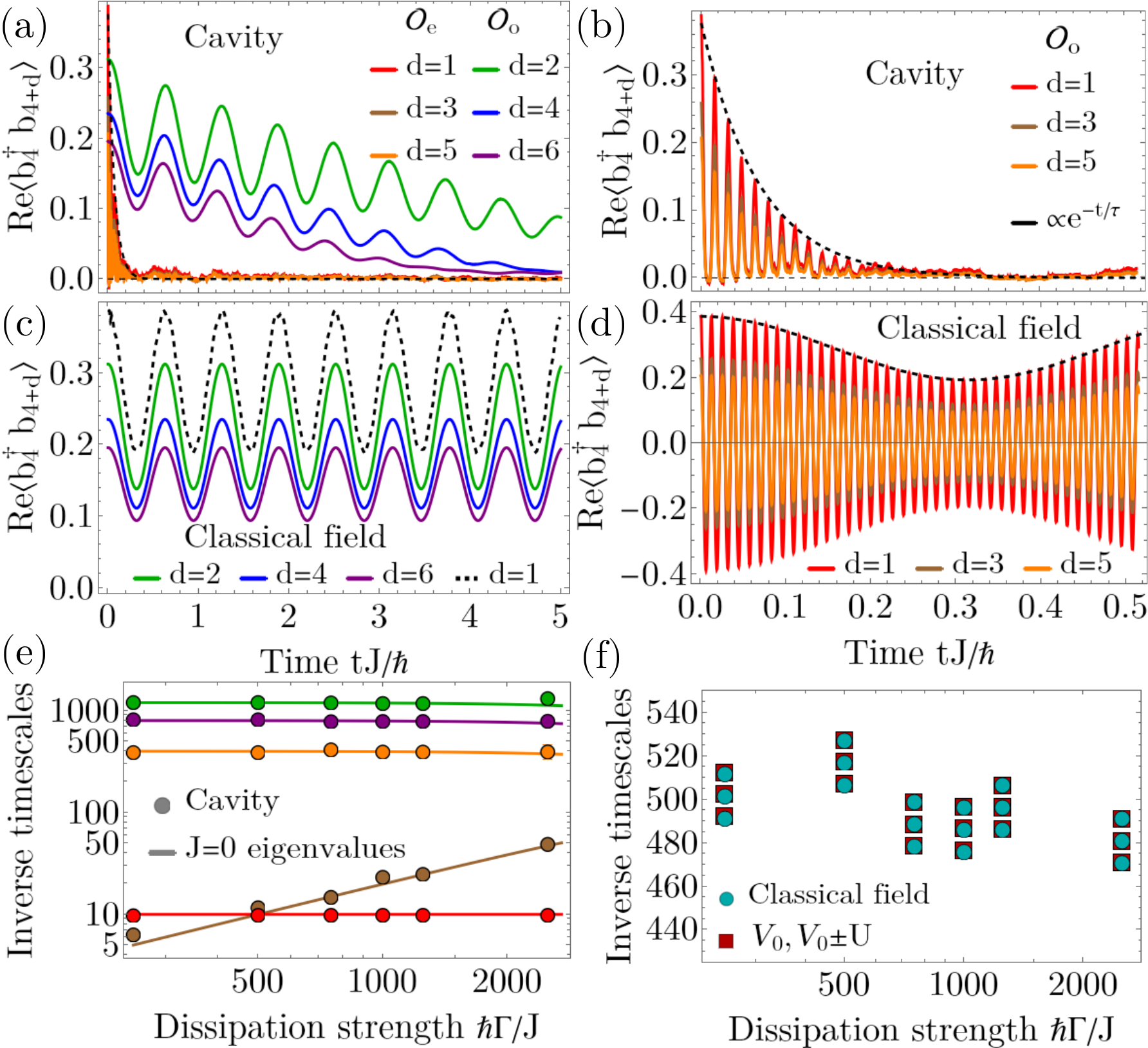}
\caption{Time evolution of the single particle correlations 
$\Re\left\langle b^\dagger_4 b_{4+d} \right\rangle$, (a)-(b) the exact description of the cavity, Eq.~(\ref{eq:Lindblad}), and (c)-(d) classical field approach, for $\hbar\Omega\sqrt{N}/J=1323$, $\hbar\delta/J=5000$, $\hbar\Gamma/J=750$, $U/J=10$, $N=7$, $L=14$.  
The dashed black curve in (a),(b) represents an exponential fit of the decay of the maxima for $\Re\left\langle b^\dagger_4 b_{5} \right\rangle$. The dashed black curve in (c),(d) represents the interpolated behavior of the maxima of $\Re\left\langle b^\dagger_4 b_{5} \right\rangle$ for the classical field case.
(e) Inverse timescales for dissipative quantum dynamics, the data is extracted from the tMPS evolution and the lines are the $J=0$ eigenvalues \cite{supp}, with red $|\Im\lambda_0|=U$, brown $|\Re\lambda_1|=\frac{2\hbar\Omega^2 \Gamma}{\delta^2+\Gamma^2/4}$, and $|\Im\lambda_1| =\frac{4\hbar\Omega^2 \delta}{\delta^2+\Gamma^2/4}(1-\Delta)$ in green ($\Delta=7$), purple ($\Delta=5$) and orange ($\Delta=3$).  
(f) Inverse timescales for classical field dynamics, extracted from the numerical simulations with circles and the late time value of the potential $V_0=V(t\approx 5)$ and $V_0\pm U$ with squares.
}
\label{fig:corr}
\end{figure}

\begin{figure}[!hbtp]
\centering
\includegraphics[width=0.48\textwidth]{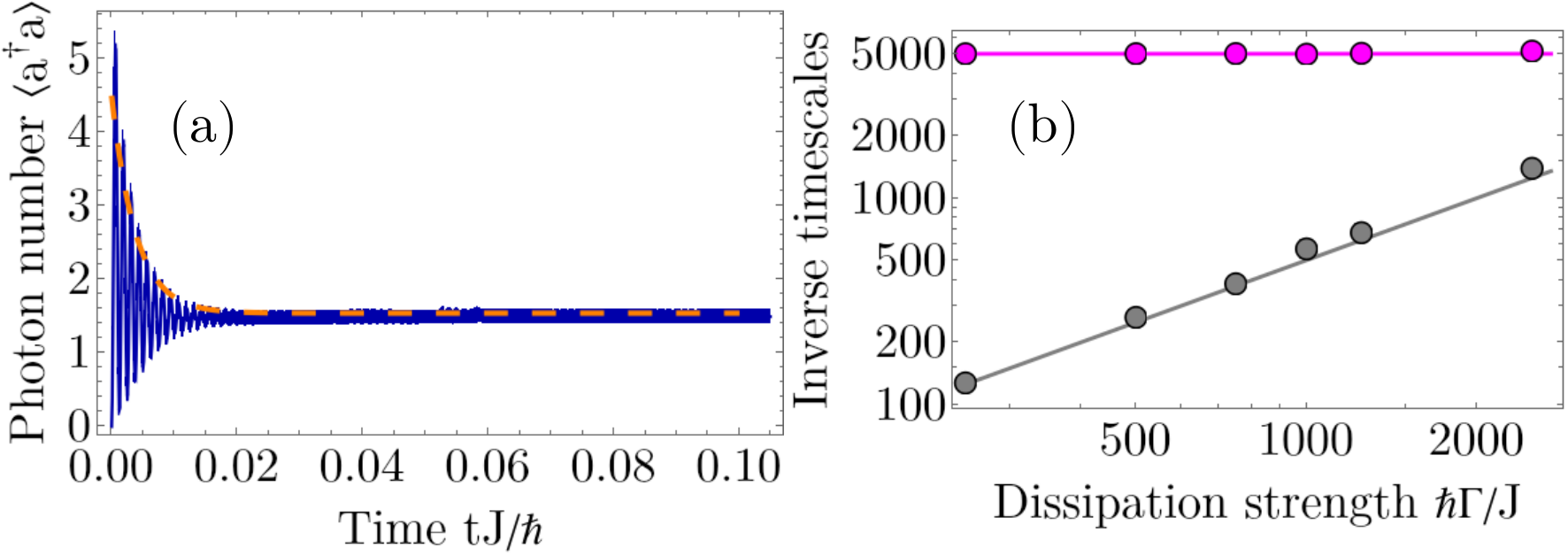}
\caption{
(a) Time evolution of the photon number, Eq.~(\ref{eq:Lindblad}).
Dashed orange line corresponds to an exponential fit of the decay of the short time oscillations, with a decay rate $\tau^{-1}/J=261\pm12\approx\hbar\Gamma/2J$.
(b) The frequency (magenta) and decay rate (gray) of the short time oscillations of the photon number versus $\hbar\Gamma/J$, the points correspond to the tMPS simulations and the lines are given by $|\Re\lambda_{P_1}|=\hbar\Gamma/2$ and $|\Im\lambda_{P_1}|=\hbar\delta$.
Parameters used are $L=14$, $N=7$, $\hbar\delta/J=5000$, $U/J=10$, (a) $\hbar\Omega\sqrt{N}/J=6614$, $\hbar\Gamma/J=500$, (b) $\hbar\Omega\sqrt{N}/J=1323$.
}
\label{fig:photons}
\end{figure}

We observe that a finite $J$, smaller than the $J=0$ gap between $\Lambda_1$ and $\Lambda_0$, induces a finite real part to all eigenvalues, except one, lying in $\Lambda_0$, Fig.~\ref{fig:sketch_ED}(d). This marks the transition from multiple steady states due the strong symmetry to a single steady state in absence of the symmetry.  
The slight breaking of the symmetry creates a subspace of long-lived metastable states only weakly coupled to dissipation, which dominate the long-time dynamics, as seen in the time-evolution of the atomic correlations, Fig.~\ref{fig:corr}(a)-(b) [same parameters as Fig.~\ref{fig:sketch_ED}(b)-(e)].
We depict the time-evolution of the single particle correlations, $\Re\left\langle b^\dagger_4 b_{4+d} \right\rangle$, obtained with the tMPS approach of simulating Eq.~(\ref{eq:Lindblad}), for a larger system \cite{supp}. 
For odd distances $d$ the correlations probe the evolution of the states contained in $\Lambda_1$, while for even distances $d$ they probe the states in the subspace $\Lambda_0$.
We observe extremely different timescales for odd and even correlations, reproducing very well the dynamics we aimed to engineer and characterized in terms of the approximate strong symmetry.
At even distances the single particle correlations show oscillations [Fig.~\ref{fig:corr}(a)], whose frequencies are determined by the value of $U$ [red points and line in Fig.~\ref{fig:corr}(e)]. The oscillations are only weakly damped on the tunneling timescale $J$. 
In contrast, for odd distances both the frequencies of the oscillations and their exponential decay to a small value occur on much faster timescales [Fig.~\ref{fig:corr}(a)-(b)]. 
We extract these timescales and obtain very good agreement with the analytical eigenvalues of $\Lambda_1$ in Fig.~\ref{fig:corr}(e) (brown for the decay rate and green, purple and orange for the frequencies).
We note that at $J=0$ the synchronized oscillations are related to the fact that the operators $b^\dagger_i b_{i+d}$, for $d$ even, can be used to construct the eigenstates with purely imaginary eigenvalues \cite{BucaJaksch2019b,BucaJaksch2022,supp}.
Similar dynamical behavior can also be observed in evolution of the pair correlations \cite{supp}.

We highlight the importance of the dissipative quantum nature of the cavity by comparing with the case of a classical field realizing a superlattice potential $H_\text{MF}$ [Fig.~\ref{fig:corr}(a)-(b) in contrast to Fig.~\ref{fig:corr}(c)-(d)].
For the single particle correlations at even distances the oscillation frequency is the same for the quantum and classical cases, given by $U$, but for the classical potential the oscillations do not show an attenuation for the times shown [Fig.~\ref{fig:corr}(c)].
For odd distances the difference in behavior is even more striking, for the classical field the frequencies of the oscillations are given by the height of the potential and on-site interactions [see Fig.~\ref{fig:corr}(d)-(e)], and the oscillations do not decay up to long times [dashed black line in Fig.~\ref{fig:corr}(c)].
Thus, the suppression of the correlations at odd distances is due to the open quantum nature of the cavity and cannot be explained at a mean-field level by a classical superlattice potential.
We note that the dynamical behavior cannot be recovered even if one adds stochastic noise in the dynamics of the classical coherent field \cite{supp}.

\emph{Photon number dynamics:}
An interesting question is which timescales are reflected in the relaxation of the photon number. 
We observe in Fig.~\ref{fig:photons}(a) that after an initial fast increase the photon number exhibits damped oscillations followed by a plateau. The oscillations frequency is consistent with the value of $\delta$ and the fast decay with the inverse timescale of $\Gamma/2$ [see Fig.~\ref{fig:photons}(b) and \cite{supp}], corresponding to $P_1$  [Fig.~\ref{fig:sketch_ED}(b)].
We note that the photon number has not reached the steady state for the latest time shown, Fig.~\ref{fig:photons}(a), the long time dynamics corresponds to timescales set by the subspace $\Lambda_0$.
Additional information is obtained by investigating the single quantum trajectories sampled in our numerical method.
The photon number indicates that the trajectories are projected quickly to subspaces spanned by states with the same imbalance $\Delta$  \cite{supp}. These results can be interpreted in connection with the phenomenon of dissipative freezing for the case of an approximate strong symmetry \cite{MunozPorras2019, HalatiKollath2022, TindallMunoz2023}.

\begin{figure}[!hbtp]
\centering
\includegraphics[width=0.48\textwidth]{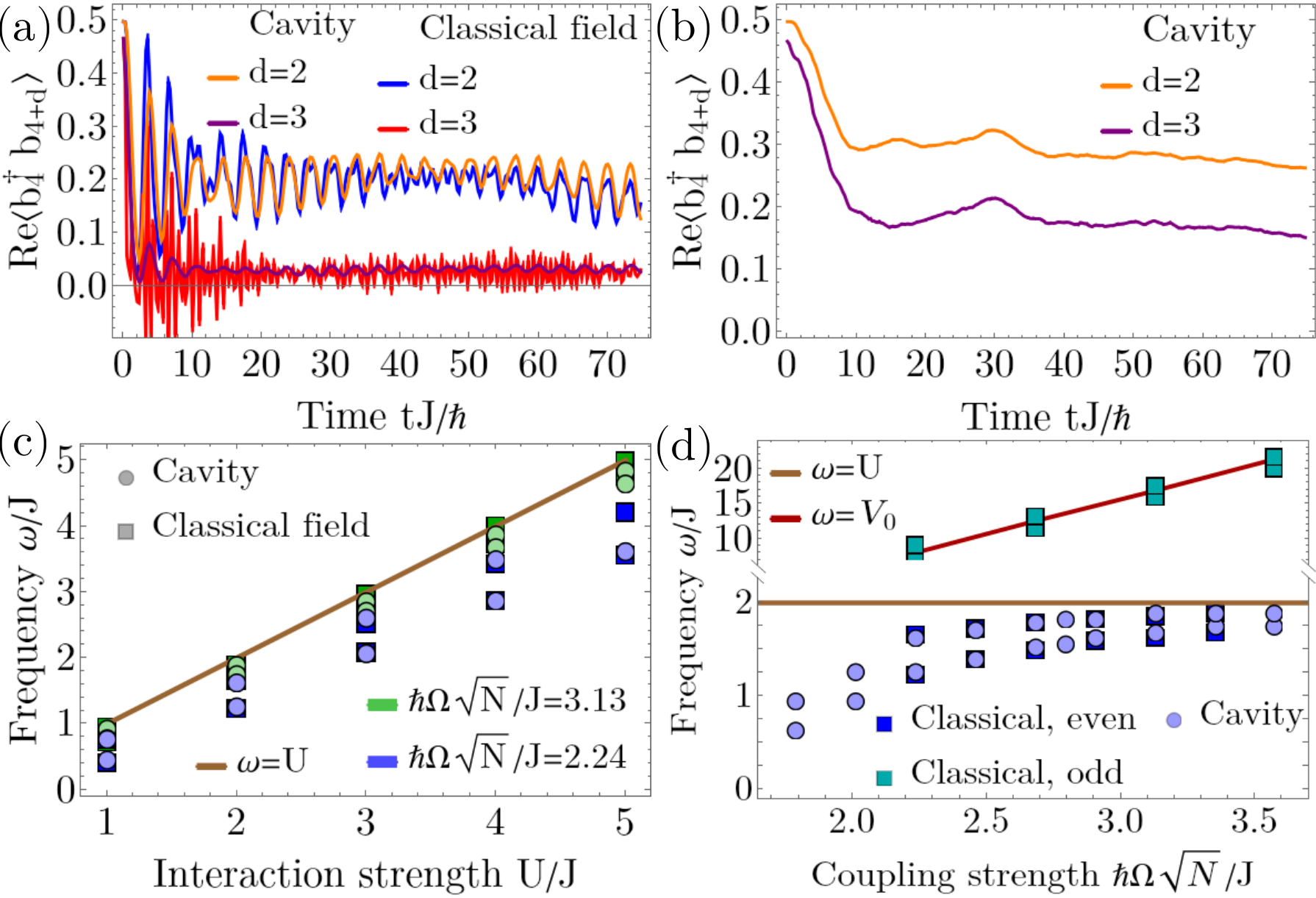}
\caption{
(a)-(b) Time evolution of the single particle correlations $\Re\left\langle b^\dagger_4 b_{4+d} \right\rangle$  (a) for a strong coupling $\hbar\Omega\sqrt{N}/J=3.35$, (b) for weak $\hbar\Omega\sqrt{N}/J=1.72$ and $U/J=2$, $L=14$, $N=7$, $\hbar\Gamma/J=1$, $\hbar\delta/J=2$
(c)-(d) Frequencies extracted from the dynamics of correlations as a function of $U$ and $\Omega$.
The lines at $\omega=U$ (brown) and $\omega=V_0$ (dark red) represent the expectation for the collapse and revival dynamics for a deep superlattice, where $V_0=V(t\approx75)$.
}
\label{fig:corr_low_dissipation}
\end{figure}

\emph{Cavity-induced self-organized synchronization:}
So far we made the connection between the timescales observed in the single particle correlations and the eigenvalues of the Liouvillian for small $J$ in the regime of large detuning and dissipation. 
Next, we show that even in regimes initially far from the strong symmetry, due to the self-organization of the cavity-atom system, an approximate symmetry arises, protecting synchronized long-lived coherences. 
To show this, we consider the very challenging regime where all parameters are comparable, see Fig.~\ref{fig:corr_low_dissipation}. In this situation, it is much harder to obtain analytical insights or track individual eigenvalues in the spectrum, however, the tMPS method allows for simulations up to long times.

At strong atoms-cavity coupling, deep in the self-organized phase \cite{HalatiKollath2020, BezvershenkoRosch2021}, we observe very similar synchronized oscillations in the atomic correlations at even distances as before, surviving up to very long times, Fig.~\ref{fig:corr_low_dissipation}(a). 
In contrast, for a coupling close to the self-organization threshold the oscillations are absent, Fig.~\ref{fig:corr_low_dissipation}(b).
In order to verify that the oscillations occurring in this regime are induced by coherences between states with different interaction energies, we compute the scaling of their frequency with $U$, Fig.~\ref{fig:corr_low_dissipation}(b). 
We obtain the linear scaling with $U$ as for the ED results in Fig.~\ref{fig:sketch_ED}(e) in the regime of small $J$.
Furthermore, going deeper into the self-organized phase by increasing $\hbar\Omega\sqrt{N}/J$, the higher value of the two frequencies approaches the value $U$, see Fig.~\ref{fig:corr_low_dissipation}(d). 
This implies that the states with coherences between configurations with the same imbalance, but different interaction energies, are long lived metastable states, producing the oscillations observed. 
For large $\Omega$ the atoms feel a strong self-organized potential, suppressing the atomic tunneling and giving rise to an emergent approximate strong symmetry, similarly to small $J$ regime. 
In contrast to the small $J$ case, in this situation a similar synchronized oscillatory behavior and dependence of the frequencies is recovered from the simulations in a classical potential [Fig.~\ref{fig:corr_low_dissipation}(a),(c),(d)] and the correlations at odd distances are not suppressed to such a small value as before, Fig.~\ref{fig:corr_low_dissipation}(a). 
After an initial decay the correlations saturate to a finite value comparable to the value obtained in the classical potential. However, in the classical potential the correlations at odd distances exhibit oscillations induced by the height of the potential [Fig.~\ref{fig:corr_low_dissipation}(d) upper part], not present for the coupling to the cavity.

\emph{Conclusions:}
We investigated how the dynamical properties of interacting atoms can be controlled by the coupling to the quantum field of a dissipative cavity.
We show that by engineering the coupling to the cavity the dynamics of atomic correlations strongly depends on the distance between the sites they probe. 
In particular, for the single particle correlations at even distances we recover a dissipative analog of the collapse and revival behavior, exhibiting metastable synchronization, i.e.~oscillatory evolution up to long times, with the frequency set by the atomic interactions. 
In contrast, the coherences at odd distances are strongly suppressed on short times, with the timescales set by the cavity parameters and atoms-cavity coupling strength.
Important insights are obtain by considering the approximate strong symmetries of the open atoms-cavity system. 
The suppression of the odd correlations stems from the fact that they probe subspaces of the Liovillian with large decay rates, while the dynamics of even correlations is contained close to the decoherence free subspace protected by the symmetry.
This offers the opportunity to induce non-trivial dynamical behavior in other many-body dissipative quantum systems. 
We further show that the approximate symmetry can arise dynamically in self-organized regime.
Experimentally, the synchronization dynamics of the coherences would be visible in the momentum distribution \cite{supp} obtained in time-of-flight measurements \cite{GreinerBloch2002b}. 
However, the momentum occupations have contributions from all single particle correlations, thus, to probe their very different evolution in-situ coherence measurements would be needed.

\emph{Data availability:} 
The supporting data for this article are openly available at Zenodo \cite{datazenodo}.

\emph{Acknowledgments:} We thank J.P.~Brantut, T.~Donner, T.~Giamarchi, S.~J\"ager, H. Ritsch, L.~Tolle, and C.~Waechtler for fruitful discussions.
We acknowledge support by the Swiss National Science Foundation under Division II grants 200020-188687 and 200020-219400 and by the  Deutsche Forschungsgemeinschaft  (DFG,  German  Research  Foundation) under project number 277625399 - TRR 185 (B4), project number 277146847 - CRC 1238 (C05), project number 429529648 - CRC-TRR 306 "QuCoLiMa", and under Germany’s Excellence Strategy – Cluster of Excellence Matter and Light for Quantum Computing (ML4Q) EXC 2004/1 – 390534769. 
This research was supported in part by the National Science Foundation under Grants No.~NSF PHY-1748958 and PHY-2309135.

\pagebreak

\section{Supplemental Material}

\setcounter{section}{0}
\renewcommand{\thesection}{\Alph{section}}
\setcounter{equation}{0}
\renewcommand{\theequation}{A.\arabic{equation}}

\subsection{\label{appA} Time-dependent matrix product states method for open cavity-atoms systems}

The numerically exact results for the time evolution of the Liouvillian, Eqs.~(1)-(2) in the main text, describing an one-dimensional Bose-Hubbard model coupled to the dissipative cavity, are obtained with a recent implementation of a matrix product states (MPS) method \cite{HalatiKollath2020, HalatiKollath2020b, HalatiPhd}. 
The method has been developed to perform the time-evolution of cavity-atoms coupled dissipative systems and the details regarding the implementation and benchmarks are presented in Ref.~\cite{HalatiKollath2020b}.
This approach employs the stochastic unravelling of the master equation with quantum trajectories \cite{DalibardMolmer1992, GardinerZoller1992, Daley2014} and a variant of the quasi-exact time-dependent variational matrix product state (tMPS) based on the Trotter-Suzuki decomposition of the time evolution propagator \cite{WhiteFeiguin2004, DaleyVidal2004, Schollwoeck2011} and the dynamical deformation of the MPS structure using swap gates \cite{StoudenmireWhite2010, Schollwoeck2011, WallRey2016}. Our implementation makes use of the ITensor Library \cite{FishmanStoudenmire2020}.

The convergence of our results is sufficient for at least 500 quantum trajectories in the Monte Carlo sampling. The other convergence parameters are chosen as described in the following:
for the results presented in Fig.~2 and Fig.~3 of the main text we use a maximal bond dimension of 100 states, which ensured a truncation error of at most $10^{-7}$, the time-step used was $dt J/\hbar=10^{-5}$, and the adaptive cutoff of the local Hilbert space of the photonic mode ranged between $N_\text{pho}=40$ and $N_\text{pho}=8$;
for the results presented in Fig.~4 of the main text we use a maximal bond dimension of 300 states, which ensured a truncation error of at most $10^{-7}$, the time-step used was $dt J/\hbar=0.0125$, and adaptive cutoff of the local Hilbert space of the photonic mode between $N_\text{pho}=45$ and $N_\text{pho}=15$.

\setcounter{equation}{0}
\renewcommand{\theequation}{B.\arabic{equation}}
\setcounter{figure}{0}
\renewcommand{\thefigure}{B\arabic{figure}}

\subsection{Dynamics of the Bose-Hubbard in a classical cavity field potential}

In the main text we contrast the dynamics of single-particle correlations of the Bose-Hubbard model coupled to the quantum dissipative field of an optical cavity with results of a closed system in which the interacting bosons are subject to a superlattice staggered potential.
In the latter case, the staggered potential can be derived as a mean-field description of the cavity-atoms coupling, approach in which the cavity field is described by a classical coherent field \cite{RitschEsslinger2013, MaschlerRitsch2008, NagyDomokos2008}.
Within this approximation the atoms are described by the Hamiltonian
\begin{align} 
\label{eq:Hamiltonian_MF}
&H_\text{MF}=H_{\text{int}}+H_{\text{kin}}+H_{\text{stag}} \\
&H_{\text{int}}=\frac{U}{2} \sum_{j=1}^L n_{j}(n_{j}-1),\nonumber\\
&H_{\text{kin}}=-J \sum_{j=1}^{L-1} (b_{j}^\dagger b_{j+1} + b_{j+1}^\dagger b_{j}), \nonumber\\
&H_{\text{stag}}=  -V(t) \Delta, ~~ \Delta=\sum_{j=1}^L (-1)^j n_j \nonumber,
\end{align}
where $V(t)$ mimics the coupling to a classical cavity field described by a time-dependent coherent state.
For the purpose of the comparison performed in the main text regarding the behavior of the atomic correlations, we derive $V(t)$ from exact time-dependence of the photon number via the mean-field relation 
\begin{align} 
\label{eq:MF_potential}
&V_\text{exact}(t)=\frac{2\hbar\delta\Omega}{\sqrt{\delta^2+\Gamma^2/4}}\sqrt{\left\langle a^\dagger a\right\rangle_\text{exact}(t)},
\end{align}
where $\left\langle a^\dagger a\right\rangle_\text{exact}(t)$ is given by the full model, Eqs.~(1)-(2) in the main text, for example shown in Fig.~3 of the main text.
This choice allows for a comparison between a quantum dissipative and classical potentials, which contain the same time-scales and average behavior, thus, pinpointing the role of the nature of the dissipation and fluctuation effects stemming from the cavity-atoms coupling.

As in the effective Hamiltonian given in Eq.~(\ref{eq:Hamiltonian_MF}) we impose the sign of the classical potential coupled to the atomic imbalance, we break the $\mathbb{Z}_2$ weak symmetry found in the full Liouvillian (Eqs.~(1)-(2) in the main text) associated to changing the sign of the cavity field. 
Thus, in order to be able to compare the results between the exact evolution and the mean-field approach we perform the time evolution with both signs $H_{\text{stag}}=\pm V(t) \Delta$ and average the results to describe a mixture of states with the two possible signs of the imbalance and recover the $\mathbb{Z}_2$ symmetry.

\begin{figure}[!hbtp]
\centering
\includegraphics[width=0.49\textwidth]{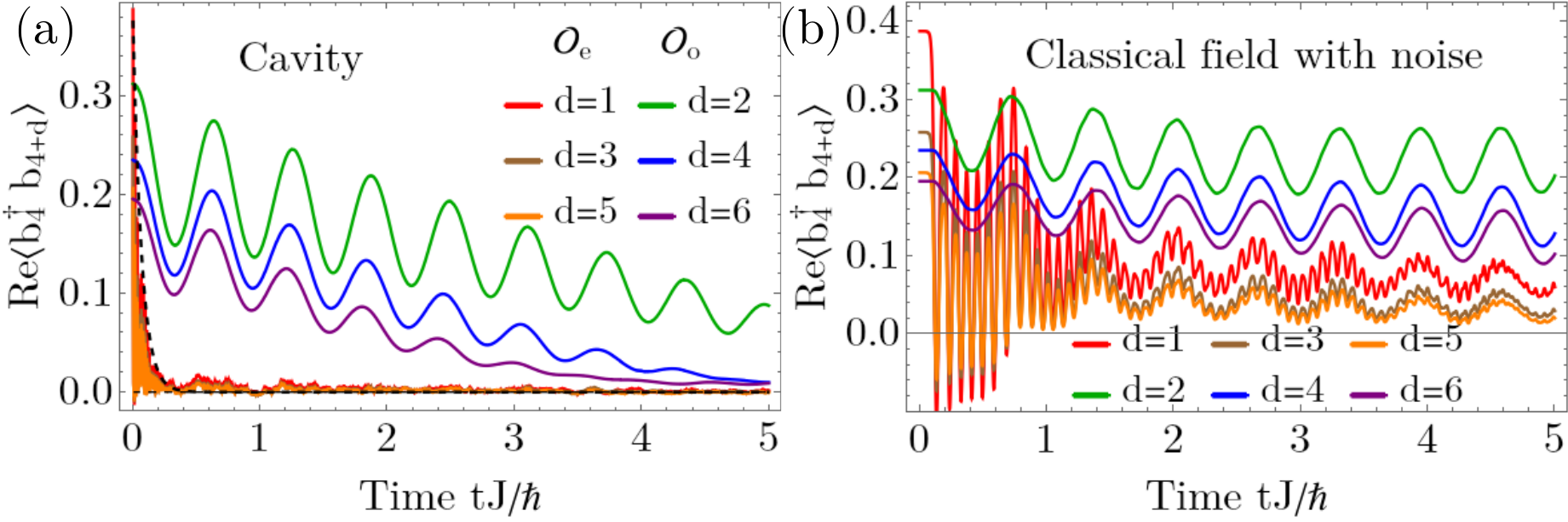}
\caption{Time evolution of the single particle correlations 
$\Re\left\langle b^\dagger_4 b_{4+d} \right\rangle$, for the (a) dissipative quantum description of the cavity, same as Fig.~2(a) of the main text, and (b) classical field with noise approach, Eq.~(\ref{eq:mf_photon_field}), for $\hbar\Omega\sqrt{N}/J=1323$, $\hbar\delta/J=5000$, $\hbar\Gamma/J=750$, $U/J=10$, $N=7$, $L=14$.  
}
\label{fig:supp_corr}
\end{figure}

To further emphasize the importance of the quantum nature of the cavity field for obtaining the dynamics of the single particle correlations highlighted in the main text, we perform a second comparison.
In this approach we consider a classical coherent cavity field described by equation of motion obtained after performing a mean-field decoupling of the atoms and the cavity
\begin{align} 
\label{eq:mf_photon_field}
&\frac{\partial}{\partial t} \langle a \rangle = i \Omega \langle\Delta\rangle-\left(i\delta+\Gamma/2\right)\langle a \rangle+\sqrt{\Gamma}\xi(t),
\end{align}
where $\xi(t)$ is a random complex number of magnitude $1$ sampled from a uniform distribution at each point in time. Furthermore, we can similarly use a uniformly sampled random complex number for the initial state $\langle a \rangle (0)$. 
We can integrate this equation of motion in a coupled way with the time-evolution of $H_\text{MF}$, where the potential coupled to the atomic odd-even imbalance is
\begin{align} 
\label{eq:sc_potential}
&V_\text{sc}(t)=\Omega \left\langle a^\dagger+ a\right\rangle(t)
\end{align}
and depends on the imbalance $\langle\Delta\rangle (t-\text{d}t)$ at the previous time step.
In practice, we simulate different realizations of the time-dependent noise term, similar to the quantum trajectories approach, and average over the values of the computed observables for the different realizations.

As in Fig.~2 of the main text in Fig.~\ref{fig:supp_corr} we compare the numerically exact behavior of the single particle correlations at different distances with the approach using a classical coherent field under the action of stochastic noise. We observe that even though the noise adds a dissipative character to the classical field it does not recover the behavior obtained for the dissipative quantum field description of the cavity.
For the correlations at even distances we obtain the oscillations with the frequency of $U$ and the noise induces a slight damping weaker compared to exact evolution [in contrast to Fig.~2(c) of the main text]. However, the correlations at odd distances still behave very differently, neither their decay timescale nor the frequency of their oscillations match.

\setcounter{equation}{0}
\renewcommand{\theequation}{C.\arabic{equation}}

\subsection{Symmetries and Liouvillian spectrum for vanishing hopping $J=0$ \label{app:sym_spectrum}}

In the first part of the main text we consider parameter regimes in which a separation of scales exists between the parameters describing the cavity field and the ones corresponding to the atoms, $\hbar \Gamma, \hbar \Omega, \hbar\delta\gg J$. Thus, it is useful to get an intuition regarding the spectrum of the Liouvillian in the limit of vanishing hopping, $J=0$,
\begin{align}
\label{eq:Lindbladfree}
&\mathcal{L}_{0}(\cdot)=-\frac{i}{\hbar}[H_\text{c}+H_{\text{int}}+H_{\text{ac}},\cdot]+\mathcal{D}(\cdot).
\end{align}
In this limit we can analytically compute the eigenstates and eigenvalues of $\mathcal{L}_{0}$. We can observe that in $\mathcal{L}_{0}$ all atomic operators are diagonal in the position basis due to a strong symmetry \cite{AlbertJiang2014, BucaProsen2012}. For an open quantum system to have a strong symmetry, the generators of the symmetry need to commute with both the Hamiltonian and the jump operators, which in our case are the local atomic density operators.

We can block diagonalize the Liouvillian into symmetry sectors of the form $\ket{pho;\{n_j\}}\bra{pho',\{n'_j\}}$, where the atomic states are fully characterized by the densities distributions $\{n_j\}$ and $\{n'_j\}$, and the photonic states $\ket{pho}$ and $\bra{pho'}$ need to be determined.
Only for the case that $\{n_j\}$ and $\{n'_j\}$ are identical, the sector contains a physical steady state, otherwise the sector only consists of traceless states.

In the following, we discuss some eigenstates of $\mathcal{L}_{0}$ and their corresponding right eigenvalues for $J=0$ in more detail. The steady states, which have an eigenvalue $\lambda_{0,\text{st}}=0$ are given by \cite{HalatiKollath2020b, SchuetzMorigi2013}
\begin{align}
\label{eq:states_steady}
& \rho_{0,st}=\ket{\alpha(\Delta);\{n_j\}}\bra{\alpha(\Delta);\{n_j\}},
\end{align}
with the atomic part diagonal in the Fock space, with the local densities $\{n_j\}$ parameterizing the symmetry sector. 
The photons are in a coherent state which depends on the atomic imbalance
$\alpha(\Delta)=\frac{\Omega}{\delta-i\Gamma/2} \Delta$, with  $\Delta=\sum_{j} (-1)^j n_j$.
One can find excited eigenstates in these sectors by creating photon excitations on top of the coherent state, for example in the subspace of single photon excitations $P_1$, we have
\begin{align}
\label{eq:states_P1}
&\rho_{P_1}=\left[a^\dagger-\alpha(\Delta)^*\right]\ket{\alpha(\Delta);\{n_j\}}\bra{\alpha(\Delta);\{n_j\}},
\end{align}
with
\begin{align}
\label{eq:eigenvalues_P1}
& \lambda_{P_1}=-\frac{\hbar\Gamma}{2}-i\hbar\delta.
\end{align}
One can show that $\rho_{P_1}$ is an eigenstate by performing a displacement of the photonic operators with the value $\alpha(\Delta)$. 
Furthermore, a harmonic oscillator ladder is obtained with the creation ladder operators given by $a^\dagger-\alpha(\Delta)^*$ generating multi-photon excitations. Thus, the eigenvalues corresponding to these multi-photon excitations will have a real part spaced by integer multiples of $-\hbar\Gamma/2$.
We note that the adjoint state $\rho_{P_1}^{\dagger}$ is also an eigenstate with the eigenvalue $\lambda_{P_1}^*$.

Next, one can show that states that are not diagonal in the atomic part, of the form $\ket{\alpha(\Delta);\{n_j\}}\bra{\alpha(\Delta');\{n'_j\}}$
are eigenstates with the eigenvalues 
\begin{align}
\label{eq:eigenvalue_all}
\lambda(\Delta,u,\Delta',u')& = - \frac{1}{2}\frac{\hbar\Omega^2\Gamma}{\delta^2+\Gamma^2/4}(\Delta-\Delta')^2 \\
&+i\left[\frac{\hbar\Omega^2\delta}{\delta^2+\Gamma^2/4}(\Delta^2-\Delta'^2)-(u-u')\right], \nonumber
\end{align}
with the odd-even imbalances, $\Delta=\sum_{j} (-1)^j n_j$ and $\Delta'=\sum_{j} (-1)^j n_j'$ and the total interaction energies $u= \frac{U}{2}\sum_{j}n_j(n_j-1)$ and $u'= \frac{U}{2}\sum_{j}n_j'(n_j'-1)$.
Based on the values of the imbalances and interaction energies we can identify the following cases:
\begin{itemize}
\item For $\Delta=\Delta'$, even for different density distributions $\{n_j\}$ and $\{n'_j\}$, the states
\begin{align}
\label{eq:states_low}
&\rho_0=\ket{\alpha(\Delta);\{n_j\}}\bra{\alpha(\Delta);\{n'_j\}},
\end{align}
have eigenvalues with vanishing real part
\begin{align}
\label{eq:eigenvalue_low}
\lambda_0=-i(u-u'). 
\end{align}
In the case $u=u'$ we have traceless states with $\lambda_0=0$, showing the existence of off-diagonal coherences between Fock states that can survive in the steady states.
Alternatively, for $u\neq u'$, we obtain states with purely imaginary eigenvalues. These states, called rotating coherences \cite{AlbertJiang2014}, can lead to persistent synchronized oscillations in the long time limit \cite{BucaJaksch2019b,BucaJaksch2022}.

The evolution of some of these states can be probed by monitoring, for example, the evolution of single particle correlations at even distances $b^\dagger_j b_{j+2d}$ \cite{BucaJaksch2019b,BucaJaksch2022}, as they probe the coherences induced by states which have the same imbalance $\Delta$ in the bra and the ket contribution.

Furthermore, as shown in Ref.~\cite{BucaJaksch2019b,BucaJaksch2022}, we can use the operators $b^\dagger_i b_{i+2d}$ to construct the states with purely imaginary eigenvalues. As the following conditions are fulfilled
\begin{align}
\label{eq:eigenvalue_imaginary}
&[H_\text{c}+H_{\text{int}}+H_{\text{ac}}, b^\dagger_i b_{i+2d}]\rho_{0,st}= \\
&\qquad=-U(n_{i+2d}-1-n_i) b^\dagger_i b_{i+2d} \rho_{0,st}, \nonumber\\
&[a, b^\dagger_i b_{i+2d}]\rho_{0,st}=0, \nonumber
\end{align}
then the state $b^\dagger_j b_{j+2d}\rho_{0,st}$ is an eigenstate of the Liouvillian $\mathcal{L}_{0}$ with the eigenvalue $-U(n_{j+2d}-1-n_j)i$. Thus, we recover a subset of the states given in Eq.~(\ref{eq:states_low}).
We note that in Eq.~(\ref{eq:eigenvalue_imaginary}) we made use in the calculation that $\rho_{0,st}$ is diagonal in the Fock space.

All the states with a zero real part, Eq.~(\ref{eq:states_steady}) and Eq.~(\ref{eq:states_low}), are part of the decoherence free subspace $\Lambda_0$ (see Fig.~1 in the main text).

\item For $\Delta\neq\Delta'$ we observe in Eq.~(\ref{eq:eigenvalue_all}) that the real part of the eigenvalues is finite and proportional to $(\Delta-\Delta')^2$. Thus, the states with the lowest decaying rate are the ones for which $\Delta'=\Delta\pm 2$, which we mark by the subspace $\Lambda_1$ in Fig.~1 in the main text, 
\begin{align}
\label{eq:states_1}
&\rho_1=\ket{\alpha(\Delta);\{n_j\}}\bra{\alpha(\Delta\pm2);\{n'_j\}},
\end{align}
corresponding to the eigenvalues
\begin{align}
\label{eq:eigenvalue_1}
\lambda_1&(\Delta,u,\Delta\pm2,u')= - \frac{2\hbar\Omega^2\Gamma}{\delta^2+\Gamma^2/4}\\
&
-i\left[\frac{4\hbar\Omega^2\delta}{\delta^2+\Gamma^2/4}(1\pm\Delta)+(u-u')\right]. \nonumber
\end{align}
Depending on the initial state, the decay of such states can be observed in the evolution of the single particle correlations at odd distances $b^\dagger_j b_{j+2d+1}$.
\end{itemize}

We note that a similar construction as in Eq.~(\ref{eq:states_P1}) is possible to describe the photonic excitations in the symmetry sectors in which the lowest decaying states are given by the ones of Eq.~(\ref{eq:states_low}) and Eq.~(\ref{eq:states_1}).

Furthermore, with the knowledge of the different dissipative subspaces for $J=0$ one can compute perturbatively the steady state for finite and small $J$ \cite{PolettiKollath2013, SciollaKollath2015, HalatiKollath2020b}. Considering the contributions from the subspaces that can be accessed via one hopping event, $\Lambda_1$ and back, the effective dynamics for the elements of the decoherence free subspace is given by
\begin{align}
\label{eq:decfree0_kin}
\pdv{t} \rho_{0} = \lambda_0 \rho_0+\frac{1}{\hbar^2} X_0 \left[ H_{\text{kin}},\mathcal{L}_{0}^{-1} X_1 \left[H_{\text{kin}},\rho_0\right]\right].
\end{align}
The operators $X_0$ and $X_1$ are the projectors onto the decoherence free subspace $\Lambda_0$ and excited subspace $\Lambda_1$.
The kinetic term breaks the strong symmetry of $\mathcal{L}_{0}$, which determines a transition from multiple steady states to a unique steady state given by the mixed state \cite{HalatiKollath2020b, HalatiPhd}
\begin{align}
\label{eq:ss_kin}
\rho_{\text{mix}} &=\frac{1}{\mathcal{N}}\sum_{\{n_j\}} \ket{\alpha(\Delta);\{n_j\}}\bra{\alpha(\Delta);\{n_j\}},
\end{align}
where the sum runs over all possible density configurations $\{n_j\}$ and $\mathcal{N}$ is the number of these configurations. 
The state $\rho_{\text{mix}}$ exhibits strong, however classical, correlations between the cavity field and the atoms, as for each term in the sum the cavity field is fully determined by the atomic imbalance. By tracing out the photonic states we obtain a fully mixed, infinite temperature, state for the atoms as all density configurations have the same probability.

\setcounter{equation}{0}
\renewcommand{\theequation}{D.\arabic{equation}}
\setcounter{figure}{0}
\renewcommand{\thefigure}{D\arabic{figure}}

\subsection{Dynamics of single quantum trajectories}

\begin{figure}[!hbtp]
\centering
\includegraphics[width=0.48\textwidth]{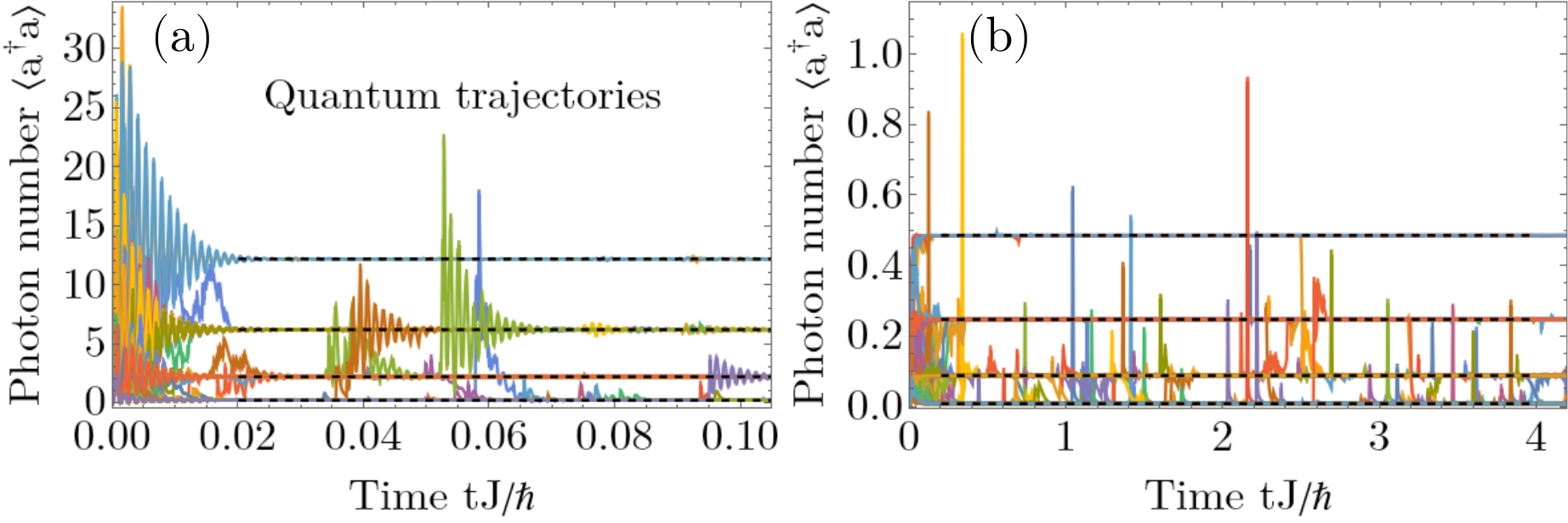}
\caption{
Time evolution of the photon number for 500 of the sampled quantum trajectories, for the parameters
$L=14$, $N=7$ , $\hbar\delta/J=5000$, $U/J=10$, (a) $\hbar\Omega\sqrt{N}/J=6614$, $\hbar\Gamma/J=500$, (b) $\hbar\Omega\sqrt{N}/J=1323$, $\hbar\Gamma/J=750$.
The dashed black lines represent the photon number expected for the possible values of the imbalance $\Delta\in\{\pm 1,\pm 3,\pm 5,\pm 7\}$.
}
\label{fig:trajectories_photons}
\end{figure}

\begin{figure}[!hbtp]
\centering
\includegraphics[width=0.48\textwidth]{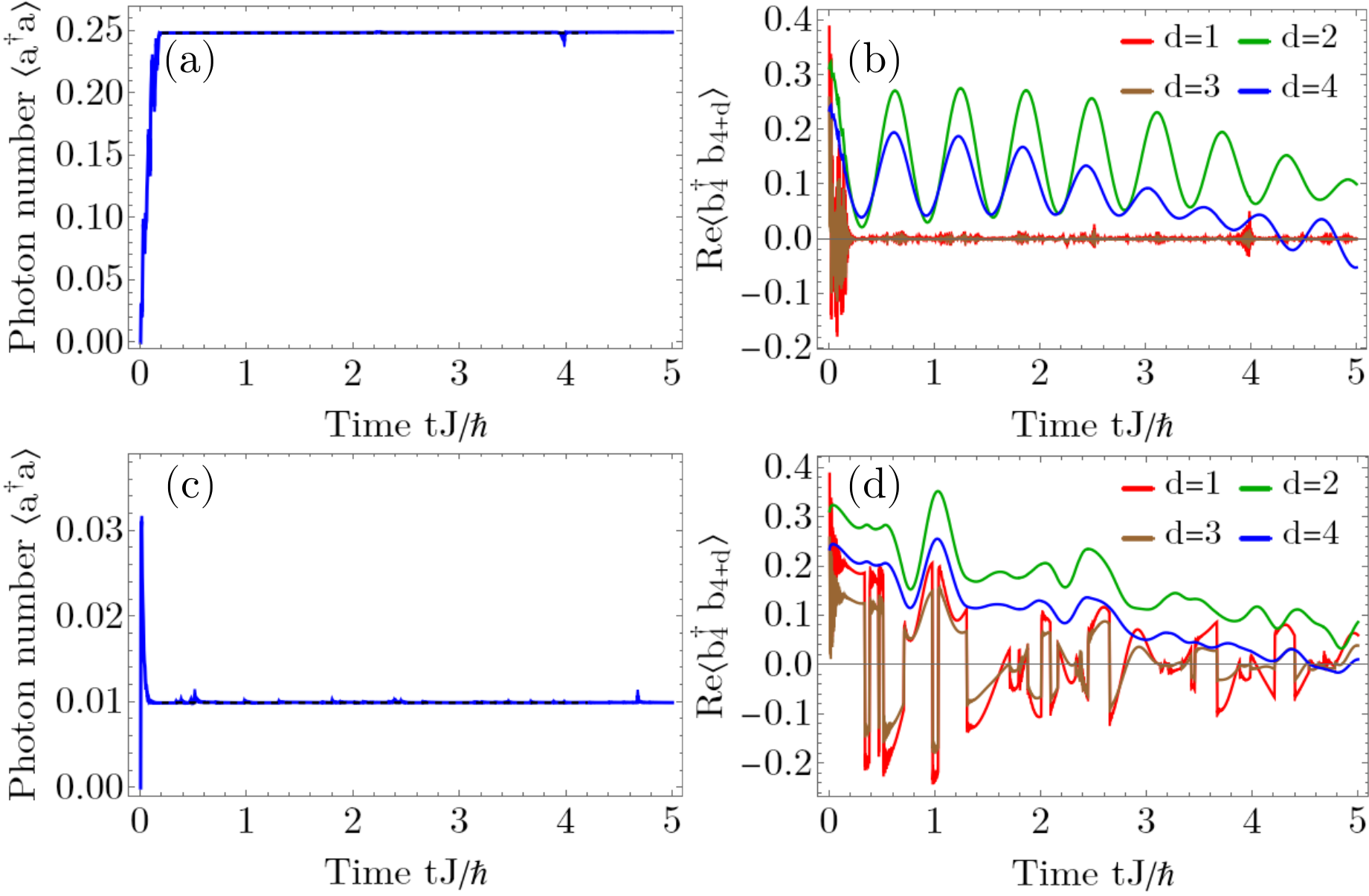}
\caption{
Time evolution of single quantum trajectories for (a), (c) the photon number and (b), (d) single particle correlations $\Re\left\langle b^\dagger_4 b_{4+d} \right\rangle$. Panels (a)-(b) correspond to the same trajectory from the subspace $\Delta=\pm5$, while panels (c)-(d) correspond to a trajectory from the subspace $\Delta=\pm1$.
The parameters used are
$L=14$, $N=7$, $\hbar\delta/J=5000$, $U/J=10$, $\hbar\Omega\sqrt{N}/J=1323$, $\hbar\Gamma/J=750$.
}
\label{fig:trajectories_corr}
\end{figure}

In the presence of a strong symmetry the quantum trajectories obtained in the stochastic unraveling of the master equation can exhibit dissipative freezing \cite{MunozPorras2019, HalatiKollath2022, TindallMunoz2023}, i.e.~the quantum trajectories can dynamically break the strong symmetry and be projected to just one of the symmetry sectors. This phenomenon occurs when the initial state consists of a superposition of states from different symmetry sectors. The evolution can be sketched as in the following
\begin{align}
\label{eq:diss_freeze}
&\ket{\psi_k(t=0)}=\sum_{\{n_j\}} c\left[\{n_j\}\right] \ket{\{n_j\}} \xrightarrow[]{\text{projected}}\\
& \ket{\psi_k (t\gg 0)}=\ket{\{n_j\}}~\text{with probability}~|c\left[\{n_j\}\right]|^2, \nonumber
\end{align}
where $\{n_j\}$ represents the set of conserved quantities of the symmetry, in our case the local densities, and $c\left[\{n_j\}\right]$ the initial amplitudes for each state.
We note that in the case in which multiple states with the same values of the conserved quantities exist, the state $\ket{\psi_k (t)}$ can still evolve in time within the symmetry sector.
This evolution implies that the steady state of the system is described by a mixed state of the form $\rho_\text{mix}=\sum |c\left[\{n_j\}\right]|^2 \ket{\{n_j\}}\bra{\{n_j\}}$, without any coherences between states from different symmetry sectors.
Thus, a necessary condition for dissipative freezing to occur is that no traceless eigenstates with $0$ real part of their eigenvalues which correspond to coherences of different symmetry sectors are present \cite{TindallMunoz2023}.
Another exception exists when one has similar symmetry sectors, for which one can find an unitary transformation to map the Hamiltonian of one sector to the Hamiltonian of the other, while only changing the jump operators up to a phase factor \cite{TindallMunoz2023}.

In the case of the system considered here [Eqs. (1)-(2) in the main text] we have shown in the previous section, Eqs.~(\ref{eq:states_low})-(\ref{eq:eigenvalue_low}), that in the limit $J=0$ states corresponding to coherences between different density configurations, and thus different symmetry sectors, are present in the steady state. Furthermore, we also have similar symmetry sectors for which $n_j=n_{L+1-j}'$ for all $j=1,\dots,L$, These sectors have an imbalance of opposite signs and can be mapped to each other by the transformation which changes the sign of the jump operator $a\to -a$.
Thus, it is interesting to observe in the dynamics of quantum trajectories if dissipative freezing can still occur between certain symmetry sectors, while the coherences between others survive to long times.

In Fig.~\ref{fig:trajectories_photons} we show the photon number corresponding to single quantum trajectories in the case of small hopping $J$, which slightly breaks the strong symmetry.
However, we have previously shown that also in the presence of an approximate strong symmetry \cite{HalatiKollath2022}, one can still interpret the results in the context of dissipative freezing, as the quantum trajectories are initially projected to the symmetry sectors and only on longer timescales, given by the symmetry breaking term, explore other subspaces.
We observe that on short times, $tJ/\hbar\sim 0.02$ in Fig.~\ref{fig:trajectories_photons}(a) and $tJ/\hbar\sim 0.1$ in Fig.~\ref{fig:trajectories_photons}(b) all 500 trajectories stabilized to a value of the photon number corresponding to one of the possible values of the imbalance $\Delta\in\{\pm 1,\pm 3,\pm 5,\pm 7\}$.
This implies that the quantum trajectories are projected to subspaces spanned by states with the same absolute value of the imbalance.
Furthermore by investigating the behavior of the single particle correlations for a trajectory, e.g.~corresponding to an imbalance $|\Delta|=5$ in Fig.~\ref{fig:trajectories_corr}(b), we can infer that coherences within these subspaces are maintained. 
We can see that at even distances $\Re\left\langle b^\dagger_4 b_{4+d} \right\rangle$ exhibits the oscillating dynamics we observed in the Monte Carlo average and discussed in the main text, showing the presence of coherences within the single trajectory. This in contrast to correlations at odd distance which are quickly suppressed and quantify the coherence between sectors with a different $|\Delta|$.
The behavior of the odd distance single particle correlations is different for the quantum trajectories which are projected to the symmetry sectors with $\Delta=\pm 1$, Fig.~\ref{fig:trajectories_corr}(d). As mentioned above $\Delta=1$ and $\Delta=-1$ are similar symmetry sectors and the coherence between them is not suppressed in the quantum trajectories. Thus, we see in Fig.~\ref{fig:trajectories_corr}(d) that the correlations at odd distances are finite, however they change sign every time a quantum jump occurs, which leads to a small value in the Monte Carlo average over all the trajectories.

As we consider a finite $J$ we also see in Fig.~\ref{fig:trajectories_photons} that rarely there are trajectories that change the subspace of fixed imbalance, capturing the long-time dynamics of approaching the steady state \cite{HalatiKollath2022}.

Thus, we see that the proximity of a strong symmetry of the open system is crucial for the correlation dynamics. The protection of the oscillations present in the single particle correlations at even distances stems from the long lived coherences between degenerate approximate symmetry sectors, while the suppression of the correlations at odd distances is due to the fact that they couple to coherence of distinct symmetry sectors with finite decay rate. 

\setcounter{equation}{0}
\renewcommand{\theequation}{E.\arabic{equation}}
\setcounter{figure}{0}
\renewcommand{\thefigure}{E\arabic{figure}}

\subsection{Results for a different initial state \label{sec:U0_is}}

\begin{figure}[!hbtp]
\centering
\includegraphics[width=0.49\textwidth]{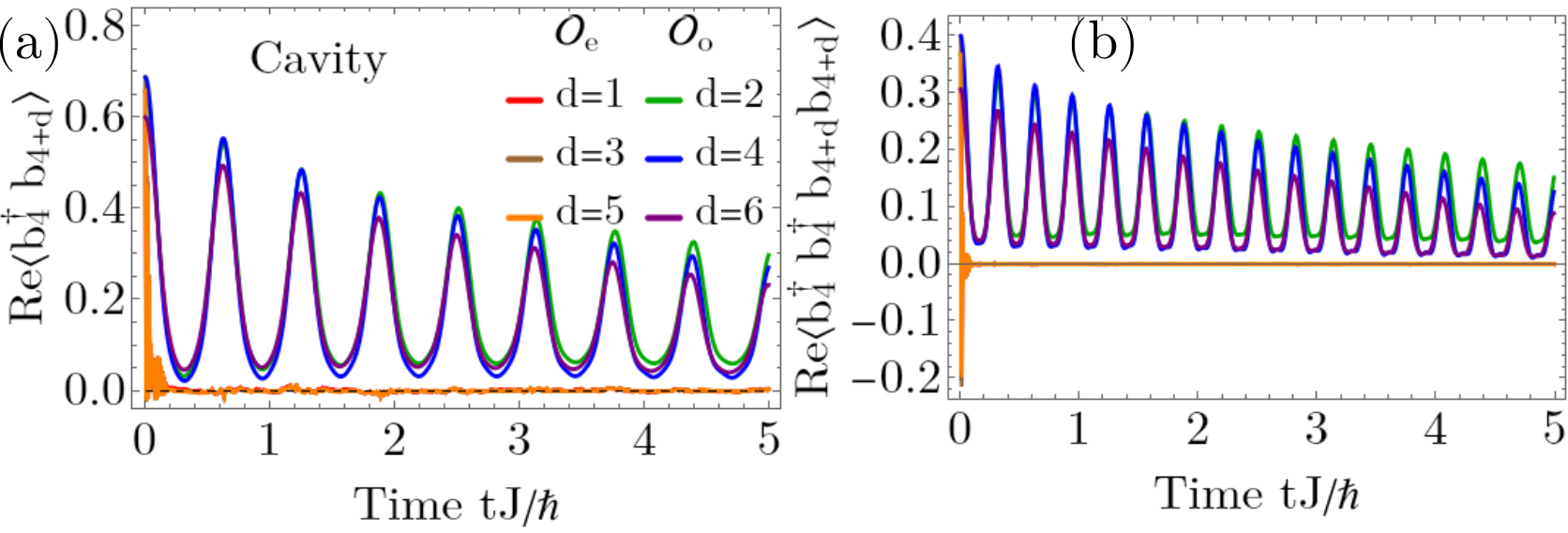}
\caption{Time evolution of (a) the single particle correlations 
$\Re\left\langle b^\dagger_4 b_{4+d} \right\rangle$, and (b) the pair correlations, $\Re\left\langle b^\dagger_4  b^\dagger_4 b_{4+d} b_{4+d} \right\rangle$, for the case in which the initial state corresponds to the ground state of the non-interacting Bose-Hubbard model for $N=7$, $L=14$, during the time-evolution the following parameters were used $\hbar\Omega\sqrt{N}/J=1323$, $\hbar\delta/J=5000$, $\hbar\Gamma/J=750$, $U/J=10$. 
}
\label{fig:supp_corr_U0}
\end{figure}

In the following, we present results for the dynamics of the single particle correlations, $\Re\left\langle b^\dagger_4 b_{4+d} \right\rangle$, and  the pair correlations, $\Re\left\langle b^\dagger_4b^\dagger_4 b_{4+d}b_{4+d} \right\rangle$, for the case in which the initial state consists of all atoms occupying the zero momentum mode, i.e.~the ground state of the non-interacting Bose-Hubbard model.
This example allows us to showcase situations in which the correlations exhibit oscillations where multiple frequency have an important contributions. The different contributions stem from states with different single site occupations resulting in different values of the interaction energies, as described in relation to eigenvalues given in Eq.~(\ref{eq:eigenvalue_low}).
In contrast, in Fig.~2 and Fig.~3 of the main text of the paper we employed as the initial state the ground state of the Bose-Hubbard model for $U/J=10$, for which the strong interactions suppress high single site occupations. 
We depict the results of the alternative initial state in Fig.~\ref{fig:supp_corr_U0}.
We can observe that as discussed in the main text, for both the single particle and the pair correlations, we get a rapid suppression of the correlations at odd distances, while the correlations at even distances are protected up to long times and exhibit an oscillatory behavior.
Compared to the results shown in the main text in Fig.~\ref{fig:supp_corr_U0}(a) for the single particle correlations at even distances if we extract the frequencies of the oscillations we obtain contributions from frequencies consistent to $\omega=U$, $\omega=2U$ and $\omega=3U$. While for the pair correlations, Fig.~\ref{fig:supp_corr_U0}(b), we obtain the main contributions from frequencies consistent with $\omega=2U$ and $\omega=4U$.

\setcounter{equation}{0}
\renewcommand{\theequation}{F.\arabic{equation}}
\setcounter{figure}{0}
\renewcommand{\thefigure}{F\arabic{figure}}

\begin{figure}[!hbtp]
\centering
\includegraphics[width=0.49\textwidth]{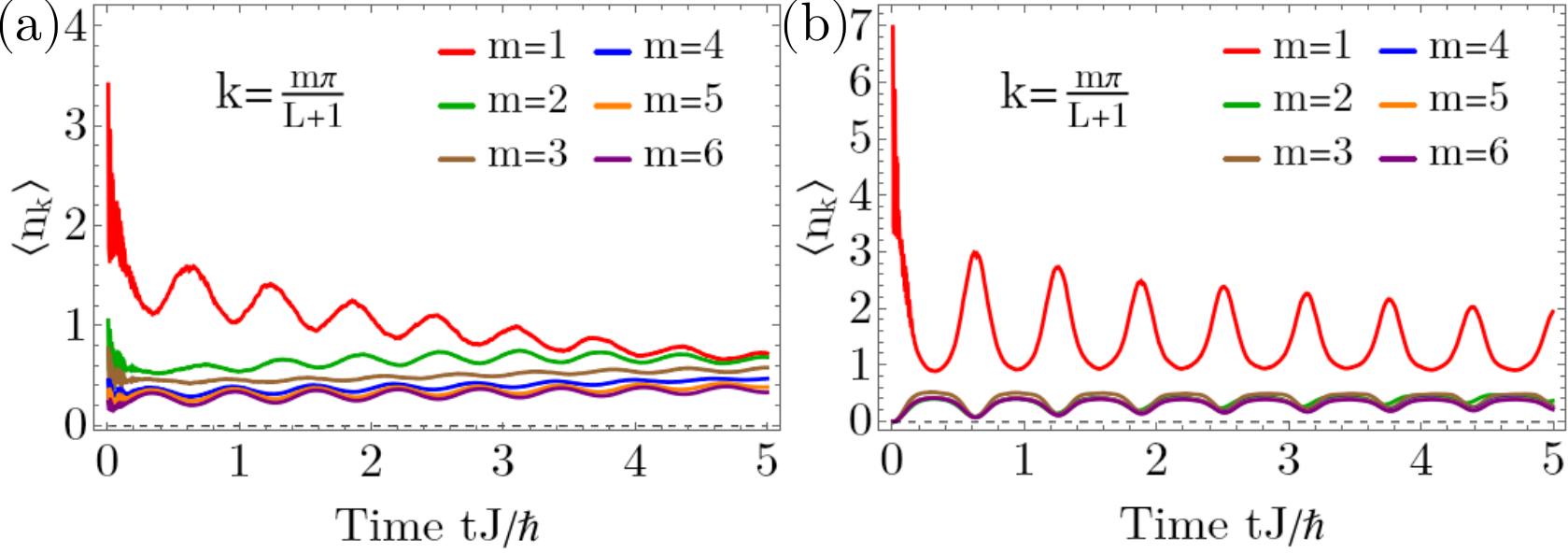}
\caption{Time evolution of the occupation of momentum states $\langle n_k\rangle=\left\langle b^\dagger_k b_{k} \right\rangle$, with $b_{k}$ defined in Eq.~(\ref{eq:op_mom}), for different values of the momentum, where the initial state is (a) ground state of the interacting Bose-Hubbard model with $U/J=10$, and (b) ground state of the non-interacting Bose-Hubbard model, during the time-evolution the following parameters were used $\hbar\Omega\sqrt{N}/J=1323$, $\hbar\delta/J=5000$, $\hbar\Gamma/J=750$, $U/J=10$, $N=7$, $L=14$. 
}
\label{fig:supp_mom}
\end{figure}

\subsection{Momentum distributions}

The metastable synchronization dynamics can be observed also in the momentum distribution, as the occupation of the different momentum states is related to the single particle correlations. As the momentum distribution is an experimentally accessible quantity we show in Fig.~\ref{fig:supp_mom} the dynamics of the occupation of the momentum modes, $\langle n_k\rangle=\left\langle b^\dagger_k b_{k} \right\rangle$, where the bosonic operators in momentum space for the finite system are defined as
\begin{align}
\label{eq:op_mom}
&b_k=\sqrt{\frac{2}{L+1}}\sum_{j=1}^L b_j \sin(kj),
\end{align}
with the unitless momenta given by $k=\frac{m\pi}{L+1}$ and $m=1,\dots,L$.
As $n_k$ probe both the $\Lambda_0$ and $\Lambda_1$ subspaces for values of momenta shown, we do not have a very different behavior of suppressing and protecting certain momentum occupations, as for the space resolved correlations.
However, the time evolution of $\langle n_k\rangle$ exhibits nicely the synchronization oscillatory dynamics, as found for the single particle correlations at even distances, for both the initial state considered in the main text [Fig.~\ref{fig:supp_mom}(a)] and in the one discussed in the previous section [Fig.~\ref{fig:supp_mom}(b)].


\begin{thebibliography}{77}%
\makeatletter
\providecommand \@ifxundefined [1]{%
 \@ifx{#1\undefined}
}%
\providecommand \@ifnum [1]{%
 \ifnum #1\expandafter \@firstoftwo
 \else \expandafter \@secondoftwo
 \fi
}%
\providecommand \@ifx [1]{%
 \ifx #1\expandafter \@firstoftwo
 \else \expandafter \@secondoftwo
 \fi
}%
\providecommand \natexlab [1]{#1}%
\providecommand \enquote  [1]{``#1''}%
\providecommand \bibnamefont  [1]{#1}%
\providecommand \bibfnamefont [1]{#1}%
\providecommand \citenamefont [1]{#1}%
\providecommand \href@noop [0]{\@secondoftwo}%
\providecommand \href [0]{\begingroup \@sanitize@url \@href}%
\providecommand \@href[1]{\@@startlink{#1}\@@href}%
\providecommand \@@href[1]{\endgroup#1\@@endlink}%
\providecommand \@sanitize@url [0]{\catcode `\\12\catcode `\$12\catcode
  `\&12\catcode `\#12\catcode `\^12\catcode `\_12\catcode `\%12\relax}%
\providecommand \@@startlink[1]{}%
\providecommand \@@endlink[0]{}%
\providecommand \url  [0]{\begingroup\@sanitize@url \@url }%
\providecommand \@url [1]{\endgroup\@href {#1}{\urlprefix }}%
\providecommand \urlprefix  [0]{URL }%
\providecommand \Eprint [0]{\href }%
\providecommand \doibase [0]{http://dx.doi.org/}%
\providecommand \selectlanguage [0]{\@gobble}%
\providecommand \bibinfo  [0]{\@secondoftwo}%
\providecommand \bibfield  [0]{\@secondoftwo}%
\providecommand \translation [1]{[#1]}%
\providecommand \BibitemOpen [0]{}%
\providecommand \bibitemStop [0]{}%
\providecommand \bibitemNoStop [0]{.\EOS\space}%
\providecommand \EOS [0]{\spacefactor3000\relax}%
\providecommand \BibitemShut  [1]{\csname bibitem#1\endcsname}%
\let\auto@bib@innerbib\@empty
\bibitem [{\citenamefont {Diehl}\ \emph {et~al.}(2008)\citenamefont {Diehl},
  \citenamefont {Micheli}, \citenamefont {Kantian}, \citenamefont {Kraus},
  \citenamefont {B{\"u}chler},\ and\ \citenamefont {Zoller}}]{DiehlZoller2008}%
  \BibitemOpen
  \bibfield  {author} {\bibinfo {author} {\bibfnamefont {S.}~\bibnamefont
  {Diehl}}, \bibinfo {author} {\bibfnamefont {A.}~\bibnamefont {Micheli}},
  \bibinfo {author} {\bibfnamefont {A.}~\bibnamefont {Kantian}}, \bibinfo
  {author} {\bibfnamefont {B.}~\bibnamefont {Kraus}}, \bibinfo {author}
  {\bibfnamefont {H.~P.}\ \bibnamefont {B{\"u}chler}}, \ and\ \bibinfo {author}
  {\bibfnamefont {P.}~\bibnamefont {Zoller}},\ }\emph {Quantum states and
  phases in driven open quantum systems with cold atoms},\ \href {\doibase
  10.1038/nphys1073} {\bibfield  {journal} {\bibinfo  {journal} {Nature
  Physics}\ }\textbf {\bibinfo {volume} {4}},\ \bibinfo {pages} {878} (\bibinfo
  {year} {2008})}\BibitemShut {NoStop}%
\bibitem [{\citenamefont {Verstraete}\ \emph {et~al.}(2009)\citenamefont
  {Verstraete}, \citenamefont {Wolf},\ and\ \citenamefont
  {Ignacio~Cirac}}]{VerstraeteCirac2009}%
  \BibitemOpen
  \bibfield  {author} {\bibinfo {author} {\bibfnamefont {F.}~\bibnamefont
  {Verstraete}}, \bibinfo {author} {\bibfnamefont {M.~M.}\ \bibnamefont
  {Wolf}}, \ and\ \bibinfo {author} {\bibfnamefont {J.}~\bibnamefont
  {Ignacio~Cirac}},\ }\emph {Quantum computation and quantum-state engineering
  driven by dissipation},\ \href {\doibase 10.1038/nphys1342} {\bibfield
  {journal} {\bibinfo  {journal} {Nature Physics}\ }\textbf {\bibinfo {volume}
  {5}},\ \bibinfo {pages} {633} (\bibinfo {year} {2009})}\BibitemShut {NoStop}%
\bibitem [{\citenamefont {Müller}\ \emph {et~al.}(2012)\citenamefont
  {Müller}, \citenamefont {Diehl}, \citenamefont {Pupillo},\ and\
  \citenamefont {Zoller}}]{MuellerZoller2012}%
  \BibitemOpen
  \bibfield  {author} {\bibinfo {author} {\bibfnamefont {M.}~\bibnamefont
  {Müller}}, \bibinfo {author} {\bibfnamefont {S.}~\bibnamefont {Diehl}},
  \bibinfo {author} {\bibfnamefont {G.}~\bibnamefont {Pupillo}}, \ and\
  \bibinfo {author} {\bibfnamefont {P.}~\bibnamefont {Zoller}},\ }\emph
  {Engineered Open Systems and Quantum Simulations with Atoms and Ions},\
  \bibfield  {booktitle} {\emph {\bibinfo {booktitle} {Advances in Atomic,
  Molecular, and Optical Physics}},\ }\href {\doibase
  https://doi.org/10.1016/B978-0-12-396482-3.00001-6} {\ \bibinfo {series}
  {Advances In Atomic, Molecular, and Optical Physics},\ \textbf {\bibinfo
  {volume} {61}},\ \bibinfo {pages} {1} (\bibinfo {year} {2012})}\BibitemShut
  {NoStop}%
\bibitem [{\citenamefont {Wiseman}\ and\ \citenamefont
  {Milburn}(2009)}]{Wiseman2009}%
  \BibitemOpen
  \bibfield  {author} {\bibinfo {author} {\bibfnamefont {H.~M.}\ \bibnamefont
  {Wiseman}}\ and\ \bibinfo {author} {\bibfnamefont {G.~J.}\ \bibnamefont
  {Milburn}},\ }\href@noop {} {\emph {\bibinfo {title} {Quantum Measurement and
  Control}}}\ (\bibinfo  {publisher} {Cambridge University Press},\ \bibinfo
  {year} {2009})\BibitemShut {NoStop}%
\bibitem [{\citenamefont {Zhang}\ \emph {et~al.}(2017)\citenamefont {Zhang},
  \citenamefont {xi~Liu}, \citenamefont {Wu}, \citenamefont {Jacobs},\ and\
  \citenamefont {Nori}}]{ZhangNori2017}%
  \BibitemOpen
  \bibfield  {author} {\bibinfo {author} {\bibfnamefont {J.}~\bibnamefont
  {Zhang}}, \bibinfo {author} {\bibfnamefont {Y.}~\bibnamefont {xi~Liu}},
  \bibinfo {author} {\bibfnamefont {R.-B.}\ \bibnamefont {Wu}}, \bibinfo
  {author} {\bibfnamefont {K.}~\bibnamefont {Jacobs}}, \ and\ \bibinfo {author}
  {\bibfnamefont {F.}~\bibnamefont {Nori}},\ }\emph {Quantum feedback: Theory,
  experiments, and applications},\ \href {\doibase
  https://doi.org/10.1016/j.physrep.2017.02.003} {\bibfield  {journal}
  {\bibinfo  {journal} {Physics Reports}\ }\textbf {\bibinfo {volume} {679}},\
  \bibinfo {pages} {1} (\bibinfo {year} {2017})}\BibitemShut {NoStop}%
\bibitem [{\citenamefont {Puente}\ \emph {et~al.}(2024)\citenamefont {Puente},
  \citenamefont {Motzoi}, \citenamefont {Calarco}, \citenamefont {Morigi},\
  and\ \citenamefont {Rizzi}}]{PuenteRizzi2024}%
  \BibitemOpen
  \bibfield  {author} {\bibinfo {author} {\bibfnamefont {D.~A.}\ \bibnamefont
  {Puente}}, \bibinfo {author} {\bibfnamefont {F.}~\bibnamefont {Motzoi}},
  \bibinfo {author} {\bibfnamefont {T.}~\bibnamefont {Calarco}}, \bibinfo
  {author} {\bibfnamefont {G.}~\bibnamefont {Morigi}}, \ and\ \bibinfo {author}
  {\bibfnamefont {M.}~\bibnamefont {Rizzi}},\ }\emph {Quantum state preparation
  via engineered ancilla resetting},\ \href {\doibase
  10.22331/q-2024-03-27-1299} {\bibfield  {journal} {\bibinfo  {journal}
  {{Quantum}}\ }\textbf {\bibinfo {volume} {8}},\ \bibinfo {pages} {1299}
  (\bibinfo {year} {2024})}\BibitemShut {NoStop}%
\bibitem [{\citenamefont {Bardyn}\ \emph {et~al.}(2013)\citenamefont {Bardyn},
  \citenamefont {Baranov}, \citenamefont {Kraus}, \citenamefont {Rico},
  \citenamefont {İmamoğlu}, \citenamefont {Zoller},\ and\ \citenamefont
  {Diehl}}]{BardynDiehl2013}%
  \BibitemOpen
  \bibfield  {author} {\bibinfo {author} {\bibfnamefont {C.-E.}\ \bibnamefont
  {Bardyn}}, \bibinfo {author} {\bibfnamefont {M.~A.}\ \bibnamefont {Baranov}},
  \bibinfo {author} {\bibfnamefont {C.~V.}\ \bibnamefont {Kraus}}, \bibinfo
  {author} {\bibfnamefont {E.}~\bibnamefont {Rico}}, \bibinfo {author}
  {\bibfnamefont {A.}~\bibnamefont {İmamoğlu}}, \bibinfo {author}
  {\bibfnamefont {P.}~\bibnamefont {Zoller}}, \ and\ \bibinfo {author}
  {\bibfnamefont {S.}~\bibnamefont {Diehl}},\ }\emph {Topology by
  dissipation},\ \href {\doibase 10.1088/1367-2630/15/8/085001} {\bibfield
  {journal} {\bibinfo  {journal} {New Journal of Physics}\ }\textbf {\bibinfo
  {volume} {15}},\ \bibinfo {pages} {085001} (\bibinfo {year}
  {2013})}\BibitemShut {NoStop}%
\bibitem [{\citenamefont {Landi}\ \emph {et~al.}(2022)\citenamefont {Landi},
  \citenamefont {Poletti},\ and\ \citenamefont {Schaller}}]{LandiSchaller2022}%
  \BibitemOpen
  \bibfield  {author} {\bibinfo {author} {\bibfnamefont {G.~T.}\ \bibnamefont
  {Landi}}, \bibinfo {author} {\bibfnamefont {D.}~\bibnamefont {Poletti}}, \
  and\ \bibinfo {author} {\bibfnamefont {G.}~\bibnamefont {Schaller}},\ }\emph
  {Nonequilibrium boundary-driven quantum systems: Models, methods, and
  properties},\ \href {\doibase 10.1103/RevModPhys.94.045006} {\bibfield
  {journal} {\bibinfo  {journal} {Rev. Mod. Phys.}\ }\textbf {\bibinfo {volume}
  {94}},\ \bibinfo {pages} {045006} (\bibinfo {year} {2022})}\BibitemShut
  {NoStop}%
\bibitem [{\citenamefont {Mivehvar}\ \emph {et~al.}(2019)\citenamefont
  {Mivehvar}, \citenamefont {Ritsch},\ and\ \citenamefont
  {Piazza}}]{MivehvarPiazza2019}%
  \BibitemOpen
  \bibfield  {author} {\bibinfo {author} {\bibfnamefont {F.}~\bibnamefont
  {Mivehvar}}, \bibinfo {author} {\bibfnamefont {H.}~\bibnamefont {Ritsch}}, \
  and\ \bibinfo {author} {\bibfnamefont {F.}~\bibnamefont {Piazza}},\ }\emph
  {Cavity-Quantum-Electrodynamical Toolbox for Quantum Magnetism},\ \href
  {\doibase 10.1103/PhysRevLett.122.113603} {\bibfield  {journal} {\bibinfo
  {journal} {Phys. Rev. Lett.}\ }\textbf {\bibinfo {volume} {122}},\ \bibinfo
  {pages} {113603} (\bibinfo {year} {2019})}\BibitemShut {NoStop}%
\bibitem [{\citenamefont {Ostermann}\ \emph {et~al.}(2019)\citenamefont
  {Ostermann}, \citenamefont {Lau}, \citenamefont {Ritsch},\ and\ \citenamefont
  {Mivehvar}}]{OstermannMivehvar2019}%
  \BibitemOpen
  \bibfield  {author} {\bibinfo {author} {\bibfnamefont {S.}~\bibnamefont
  {Ostermann}}, \bibinfo {author} {\bibfnamefont {H.-W.}\ \bibnamefont {Lau}},
  \bibinfo {author} {\bibfnamefont {H.}~\bibnamefont {Ritsch}}, \ and\ \bibinfo
  {author} {\bibfnamefont {F.}~\bibnamefont {Mivehvar}},\ }\emph
  {Cavity-induced emergent topological spin textures in a
  Bose{\textendash}Einstein condensate},\ \href {\doibase
  10.1088/1367-2630/aaf9e3} {\bibfield  {journal} {\bibinfo  {journal} {New
  Journal of Physics}\ }\textbf {\bibinfo {volume} {21}},\ \bibinfo {pages}
  {013029} (\bibinfo {year} {2019})}\BibitemShut {NoStop}%
\bibitem [{\citenamefont {Masalaeva}\ \emph {et~al.}(2021)\citenamefont
  {Masalaeva}, \citenamefont {Niedenzu}, \citenamefont {Mivehvar},\ and\
  \citenamefont {Ritsch}}]{MasalaevaRitsch2021}%
  \BibitemOpen
  \bibfield  {author} {\bibinfo {author} {\bibfnamefont {N.}~\bibnamefont
  {Masalaeva}}, \bibinfo {author} {\bibfnamefont {W.}~\bibnamefont {Niedenzu}},
  \bibinfo {author} {\bibfnamefont {F.}~\bibnamefont {Mivehvar}}, \ and\
  \bibinfo {author} {\bibfnamefont {H.}~\bibnamefont {Ritsch}},\ }\emph {Spin
  and density self-ordering in dynamic polarization gradients fields},\ \href
  {\doibase 10.1103/PhysRevResearch.3.013173} {\bibfield  {journal} {\bibinfo
  {journal} {Phys. Rev. Res.}\ }\textbf {\bibinfo {volume} {3}},\ \bibinfo
  {pages} {013173} (\bibinfo {year} {2021})}\BibitemShut {NoStop}%
\bibitem [{\citenamefont {Chiocchetta}\ \emph {et~al.}(2021)\citenamefont
  {Chiocchetta}, \citenamefont {Kiese}, \citenamefont {Zelle}, \citenamefont
  {Piazza},\ and\ \citenamefont {Diehl}}]{ChiocchettaDiehl2021}%
  \BibitemOpen
  \bibfield  {author} {\bibinfo {author} {\bibfnamefont {A.}~\bibnamefont
  {Chiocchetta}}, \bibinfo {author} {\bibfnamefont {D.}~\bibnamefont {Kiese}},
  \bibinfo {author} {\bibfnamefont {C.~P.}\ \bibnamefont {Zelle}}, \bibinfo
  {author} {\bibfnamefont {F.}~\bibnamefont {Piazza}}, \ and\ \bibinfo {author}
  {\bibfnamefont {S.}~\bibnamefont {Diehl}},\ }\emph {Cavity-induced quantum
  spin liquids},\ \href {\doibase 10.1038/s41467-021-26076-3} {\bibfield
  {journal} {\bibinfo  {journal} {Nature Communications}\ }\textbf {\bibinfo
  {volume} {12}},\ \bibinfo {pages} {5901} (\bibinfo {year}
  {2021})}\BibitemShut {NoStop}%
\bibitem [{\citenamefont {Uhrich}\ \emph {et~al.}(2023)\citenamefont {Uhrich},
  \citenamefont {Bandyopadhyay}, \citenamefont {Sauerwein}, \citenamefont
  {Sonner}, \citenamefont {Brantut},\ and\ \citenamefont
  {Hauke}}]{UhrichHauke2023}%
  \BibitemOpen
  \bibfield  {author} {\bibinfo {author} {\bibfnamefont {P.}~\bibnamefont
  {Uhrich}}, \bibinfo {author} {\bibfnamefont {S.}~\bibnamefont
  {Bandyopadhyay}}, \bibinfo {author} {\bibfnamefont {N.}~\bibnamefont
  {Sauerwein}}, \bibinfo {author} {\bibfnamefont {J.}~\bibnamefont {Sonner}},
  \bibinfo {author} {\bibfnamefont {J.-P.}\ \bibnamefont {Brantut}}, \ and\
  \bibinfo {author} {\bibfnamefont {P.}~\bibnamefont {Hauke}},\ }\href@noop {}
  {\emph {A cavity quantum electrodynamics implementation of the
  Sachdev--Ye--Kitaev model}} (\bibinfo {year} {2023}),\ \Eprint
  {http://arxiv.org/abs/2303.11343} {arXiv:2303.11343 [quant-ph]} \BibitemShut
  {NoStop}%
\bibitem [{\citenamefont {Kollath}\ \emph {et~al.}(2016)\citenamefont
  {Kollath}, \citenamefont {Sheikhan}, \citenamefont {Wolff},\ and\
  \citenamefont {Brennecke}}]{KollathBrennecke2016}%
  \BibitemOpen
  \bibfield  {author} {\bibinfo {author} {\bibfnamefont {C.}~\bibnamefont
  {Kollath}}, \bibinfo {author} {\bibfnamefont {A.}~\bibnamefont {Sheikhan}},
  \bibinfo {author} {\bibfnamefont {S.}~\bibnamefont {Wolff}}, \ and\ \bibinfo
  {author} {\bibfnamefont {F.}~\bibnamefont {Brennecke}},\ }\emph {Ultracold
  Fermions in a Cavity-Induced Artificial Magnetic Field},\ \href {\doibase
  10.1103/PhysRevLett.116.060401} {\bibfield  {journal} {\bibinfo  {journal}
  {Phys. Rev. Lett.}\ }\textbf {\bibinfo {volume} {116}},\ \bibinfo {pages}
  {060401} (\bibinfo {year} {2016})}\BibitemShut {NoStop}%
\bibitem [{\citenamefont {Sheikhan}\ \emph {et~al.}(2016)\citenamefont
  {Sheikhan}, \citenamefont {Brennecke},\ and\ \citenamefont
  {Kollath}}]{SheikhanKollath2016}%
  \BibitemOpen
  \bibfield  {author} {\bibinfo {author} {\bibfnamefont {A.}~\bibnamefont
  {Sheikhan}}, \bibinfo {author} {\bibfnamefont {F.}~\bibnamefont {Brennecke}},
  \ and\ \bibinfo {author} {\bibfnamefont {C.}~\bibnamefont {Kollath}},\ }\emph
  {Cavity-induced generation of nontrivial topological states in a
  two-dimensional Fermi gas},\ \href {\doibase 10.1103/PhysRevA.94.061603}
  {\bibfield  {journal} {\bibinfo  {journal} {Phys. Rev. A}\ }\textbf {\bibinfo
  {volume} {94}},\ \bibinfo {pages} {061603} (\bibinfo {year}
  {2016})}\BibitemShut {NoStop}%
\bibitem [{\citenamefont {Ballantine}\ \emph {et~al.}(2017)\citenamefont
  {Ballantine}, \citenamefont {Lev},\ and\ \citenamefont
  {Keeling}}]{BallantineKeeling2017}%
  \BibitemOpen
  \bibfield  {author} {\bibinfo {author} {\bibfnamefont {K.~E.}\ \bibnamefont
  {Ballantine}}, \bibinfo {author} {\bibfnamefont {B.~L.}\ \bibnamefont {Lev}},
  \ and\ \bibinfo {author} {\bibfnamefont {J.}~\bibnamefont {Keeling}},\ }\emph
  {{Meissner-like Effect for a Synthetic Gauge Field in Multimode Cavity
  QED}},\ \href {\doibase 10.1103/PhysRevLett.118.045302} {\bibfield  {journal}
  {\bibinfo  {journal} {Phys. Rev. Lett.}\ }\textbf {\bibinfo {volume} {118}},\
  \bibinfo {pages} {045302} (\bibinfo {year} {2017})}\BibitemShut {NoStop}%
\bibitem [{\citenamefont {Halati}\ \emph {et~al.}(2017)\citenamefont {Halati},
  \citenamefont {Sheikhan},\ and\ \citenamefont {Kollath}}]{HalatiKollath2017}%
  \BibitemOpen
  \bibfield  {author} {\bibinfo {author} {\bibfnamefont {C.-M.}\ \bibnamefont
  {Halati}}, \bibinfo {author} {\bibfnamefont {A.}~\bibnamefont {Sheikhan}}, \
  and\ \bibinfo {author} {\bibfnamefont {C.}~\bibnamefont {Kollath}},\ }\emph
  {Cavity-induced artificial gauge field in a Bose-Hubbard ladder},\ \href
  {\doibase 10.1103/PhysRevA.96.063621} {\bibfield  {journal} {\bibinfo
  {journal} {Phys. Rev. A}\ }\textbf {\bibinfo {volume} {96}},\ \bibinfo
  {pages} {063621} (\bibinfo {year} {2017})}\BibitemShut {NoStop}%
\bibitem [{\citenamefont {Mivehvar}\ \emph {et~al.}(2017)\citenamefont
  {Mivehvar}, \citenamefont {Ritsch},\ and\ \citenamefont
  {Piazza}}]{MivehvarPiazza2017}%
  \BibitemOpen
  \bibfield  {author} {\bibinfo {author} {\bibfnamefont {F.}~\bibnamefont
  {Mivehvar}}, \bibinfo {author} {\bibfnamefont {H.}~\bibnamefont {Ritsch}}, \
  and\ \bibinfo {author} {\bibfnamefont {F.}~\bibnamefont {Piazza}},\ }\emph
  {Superradiant Topological Peierls Insulator inside an Optical Cavity},\ \href
  {\doibase 10.1103/PhysRevLett.118.073602} {\bibfield  {journal} {\bibinfo
  {journal} {Phys. Rev. Lett.}\ }\textbf {\bibinfo {volume} {118}},\ \bibinfo
  {pages} {073602} (\bibinfo {year} {2017})}\BibitemShut {NoStop}%
\bibitem [{\citenamefont {Colella}\ \emph {et~al.}(2019)\citenamefont
  {Colella}, \citenamefont {Mivehvar}, \citenamefont {Piazza},\ and\
  \citenamefont {Ritsch}}]{ColellaRitsch2019}%
  \BibitemOpen
  \bibfield  {author} {\bibinfo {author} {\bibfnamefont {E.}~\bibnamefont
  {Colella}}, \bibinfo {author} {\bibfnamefont {F.}~\bibnamefont {Mivehvar}},
  \bibinfo {author} {\bibfnamefont {F.}~\bibnamefont {Piazza}}, \ and\ \bibinfo
  {author} {\bibfnamefont {H.}~\bibnamefont {Ritsch}},\ }\emph {Hofstadter
  butterfly in a cavity-induced dynamic synthetic magnetic field},\ \href
  {\doibase 10.1103/PhysRevB.100.224306} {\bibfield  {journal} {\bibinfo
  {journal} {Phys. Rev. B}\ }\textbf {\bibinfo {volume} {100}},\ \bibinfo
  {pages} {224306} (\bibinfo {year} {2019})}\BibitemShut {NoStop}%
\bibitem [{\citenamefont {Poletti}\ \emph {et~al.}(2013)\citenamefont
  {Poletti}, \citenamefont {Barmettler}, \citenamefont {Georges},\ and\
  \citenamefont {Kollath}}]{PolettiKollath2013}%
  \BibitemOpen
  \bibfield  {author} {\bibinfo {author} {\bibfnamefont {D.}~\bibnamefont
  {Poletti}}, \bibinfo {author} {\bibfnamefont {P.}~\bibnamefont {Barmettler}},
  \bibinfo {author} {\bibfnamefont {A.}~\bibnamefont {Georges}}, \ and\
  \bibinfo {author} {\bibfnamefont {C.}~\bibnamefont {Kollath}},\ }\emph
  {Emergence of Glasslike Dynamics for Dissipative and Strongly Interacting
  Bosons},\ \href {\doibase 10.1103/PhysRevLett.111.195301} {\bibfield
  {journal} {\bibinfo  {journal} {Phys. Rev. Lett.}\ }\textbf {\bibinfo
  {volume} {111}},\ \bibinfo {pages} {195301} (\bibinfo {year}
  {2013})}\BibitemShut {NoStop}%
\bibitem [{\citenamefont {Sciolla}\ \emph {et~al.}(2015)\citenamefont
  {Sciolla}, \citenamefont {Poletti},\ and\ \citenamefont
  {Kollath}}]{SciollaKollath2015}%
  \BibitemOpen
  \bibfield  {author} {\bibinfo {author} {\bibfnamefont {B.}~\bibnamefont
  {Sciolla}}, \bibinfo {author} {\bibfnamefont {D.}~\bibnamefont {Poletti}}, \
  and\ \bibinfo {author} {\bibfnamefont {C.}~\bibnamefont {Kollath}},\ }\emph
  {Two-Time Correlations Probing the Dynamics of Dissipative Many-Body Quantum
  Systems: Aging and Fast Relaxation},\ \href {\doibase
  10.1103/PhysRevLett.114.170401} {\bibfield  {journal} {\bibinfo  {journal}
  {Phys. Rev. Lett.}\ }\textbf {\bibinfo {volume} {114}},\ \bibinfo {pages}
  {170401} (\bibinfo {year} {2015})}\BibitemShut {NoStop}%
\bibitem [{\citenamefont {Macieszczak}\ \emph {et~al.}(2016)\citenamefont
  {Macieszczak}, \citenamefont {Gu\ifmmode \mbox{\c{t}}\else
  \c{t}\fi{}\ifmmode~\u{a}\else \u{a}\fi{}}, \citenamefont {Lesanovsky},\ and\
  \citenamefont {Garrahan}}]{MacieszczakGarrahan2016}%
  \BibitemOpen
  \bibfield  {author} {\bibinfo {author} {\bibfnamefont {K.}~\bibnamefont
  {Macieszczak}}, \bibinfo {author} {\bibfnamefont {M.}~\bibnamefont
  {Gu\ifmmode \mbox{\c{t}}\else \c{t}\fi{}\ifmmode~\u{a}\else \u{a}\fi{}}},
  \bibinfo {author} {\bibfnamefont {I.}~\bibnamefont {Lesanovsky}}, \ and\
  \bibinfo {author} {\bibfnamefont {J.~P.}\ \bibnamefont {Garrahan}},\ }\emph
  {Towards a Theory of Metastability in Open Quantum Dynamics},\ \href
  {\doibase 10.1103/PhysRevLett.116.240404} {\bibfield  {journal} {\bibinfo
  {journal} {Phys. Rev. Lett.}\ }\textbf {\bibinfo {volume} {116}},\ \bibinfo
  {pages} {240404} (\bibinfo {year} {2016})}\BibitemShut {NoStop}%
\bibitem [{\citenamefont {del Campo}\ and\ \citenamefont
  {Kim}(2019)}]{delCampoKim2019}%
  \BibitemOpen
  \bibfield  {author} {\bibinfo {author} {\bibfnamefont {A.}~\bibnamefont {del
  Campo}}\ and\ \bibinfo {author} {\bibfnamefont {K.}~\bibnamefont {Kim}},\
  }\emph {Focus on Shortcuts to Adiabaticity},\ \href {\doibase
  10.1088/1367-2630/ab1437} {\bibfield  {journal} {\bibinfo  {journal} {New
  Journal of Physics}\ }\textbf {\bibinfo {volume} {21}},\ \bibinfo {pages}
  {050201} (\bibinfo {year} {2019})}\BibitemShut {NoStop}%
\bibitem [{\citenamefont {King}\ \emph {et~al.}(2024)\citenamefont {King},
  \citenamefont {Giannelli}, \citenamefont {Menu}, \citenamefont {Kriel},\ and\
  \citenamefont {Morigi}}]{KingMorigi2024}%
  \BibitemOpen
  \bibfield  {author} {\bibinfo {author} {\bibfnamefont {E.~C.}\ \bibnamefont
  {King}}, \bibinfo {author} {\bibfnamefont {L.}~\bibnamefont {Giannelli}},
  \bibinfo {author} {\bibfnamefont {R.}~\bibnamefont {Menu}}, \bibinfo {author}
  {\bibfnamefont {J.~N.}\ \bibnamefont {Kriel}}, \ and\ \bibinfo {author}
  {\bibfnamefont {G.}~\bibnamefont {Morigi}},\ }\href
  {https://arxiv.org/abs/2311.11937} {\emph {Adiabatic quantum trajectories in
  engineered reservoirs}} (\bibinfo {year} {2024}),\ \Eprint
  {http://arxiv.org/abs/2311.11937} {arXiv:2311.11937 [quant-ph]} \BibitemShut
  {NoStop}%
\bibitem [{\citenamefont {Yi}\ \emph {et~al.}(2012)\citenamefont {Yi},
  \citenamefont {Diehl}, \citenamefont {Daley},\ and\ \citenamefont
  {Zoller}}]{YiZoller2012}%
  \BibitemOpen
  \bibfield  {author} {\bibinfo {author} {\bibfnamefont {W.}~\bibnamefont
  {Yi}}, \bibinfo {author} {\bibfnamefont {S.}~\bibnamefont {Diehl}}, \bibinfo
  {author} {\bibfnamefont {A.~J.}\ \bibnamefont {Daley}}, \ and\ \bibinfo
  {author} {\bibfnamefont {P.}~\bibnamefont {Zoller}},\ }\emph
  {Driven-dissipative many-body pairing states for cold fermionic atoms in an
  optical lattice},\ \href {http://stacks.iop.org/1367-2630/14/i=5/a=055002}
  {\bibfield  {journal} {\bibinfo  {journal} {New Journal of Physics}\ }\textbf
  {\bibinfo {volume} {14}},\ \bibinfo {pages} {055002} (\bibinfo {year}
  {2012})}\BibitemShut {NoStop}%
\bibitem [{\citenamefont {Albert}\ and\ \citenamefont
  {Jiang}(2014)}]{AlbertJiang2014}%
  \BibitemOpen
  \bibfield  {author} {\bibinfo {author} {\bibfnamefont {V.~V.}\ \bibnamefont
  {Albert}}\ and\ \bibinfo {author} {\bibfnamefont {L.}~\bibnamefont {Jiang}},\
  }\emph {Symmetries and conserved quantities in Lindblad master equations},\
  \href {\doibase 10.1103/PhysRevA.89.022118} {\bibfield  {journal} {\bibinfo
  {journal} {Phys. Rev. A}\ }\textbf {\bibinfo {volume} {89}},\ \bibinfo
  {pages} {022118} (\bibinfo {year} {2014})}\BibitemShut {NoStop}%
\bibitem [{\citenamefont {Bu{\v{c}}a}\ \emph {et~al.}(2019)\citenamefont
  {Bu{\v{c}}a}, \citenamefont {Tindall},\ and\ \citenamefont
  {Jaksch}}]{BucaJaksch2019b}%
  \BibitemOpen
  \bibfield  {author} {\bibinfo {author} {\bibfnamefont {B.}~\bibnamefont
  {Bu{\v{c}}a}}, \bibinfo {author} {\bibfnamefont {J.}~\bibnamefont {Tindall}},
  \ and\ \bibinfo {author} {\bibfnamefont {D.}~\bibnamefont {Jaksch}},\ }\emph
  {Non-stationary coherent quantum many-body dynamics through dissipation},\
  \href {\doibase 10.1038/s41467-019-09757-y} {\bibfield  {journal} {\bibinfo
  {journal} {Nature Communications}\ }\textbf {\bibinfo {volume} {10}},\
  \bibinfo {pages} {1730} (\bibinfo {year} {2019})}\BibitemShut {NoStop}%
\bibitem [{\citenamefont {Bu{\v{c}}a}\ \emph {et~al.}(2022)\citenamefont
  {Bu{\v{c}}a}, \citenamefont {Booker},\ and\ \citenamefont
  {Jaksch}}]{BucaJaksch2022}%
  \BibitemOpen
  \bibfield  {author} {\bibinfo {author} {\bibfnamefont {B.}~\bibnamefont
  {Bu{\v{c}}a}}, \bibinfo {author} {\bibfnamefont {C.}~\bibnamefont {Booker}},
  \ and\ \bibinfo {author} {\bibfnamefont {D.}~\bibnamefont {Jaksch}},\ }\emph
  {{Algebraic theory of quantum synchronization and limit cycles under
  dissipation}},\ \href {\doibase 10.21468/SciPostPhys.12.3.097} {\bibfield
  {journal} {\bibinfo  {journal} {SciPost Phys.}\ }\textbf {\bibinfo {volume}
  {12}},\ \bibinfo {pages} {097} (\bibinfo {year} {2022})}\BibitemShut
  {NoStop}%
\bibitem [{\citenamefont {Carmichael}(1991)}]{Carmichaelbook}%
  \BibitemOpen
  \bibfield  {author} {\bibinfo {author} {\bibfnamefont {H.}~\bibnamefont
  {Carmichael}},\ }\href@noop {} {\emph {\bibinfo {title} {An open systems
  approach to quantum optics}}}\ (\bibinfo  {publisher} {Springer Verlag},\
  \bibinfo {address} {Berlin Heidelberg},\ \bibinfo {year} {1991})\BibitemShut
  {NoStop}%
\bibitem [{\citenamefont {Breuer}\ and\ \citenamefont
  {Petruccione}(2002)}]{BreuerPetruccione2002}%
  \BibitemOpen
  \bibfield  {author} {\bibinfo {author} {\bibfnamefont {H.~P.}\ \bibnamefont
  {Breuer}}\ and\ \bibinfo {author} {\bibfnamefont {F.}~\bibnamefont
  {Petruccione}},\ }\href@noop {} {\emph {\bibinfo {title} {The theory of open
  quantum systems}}}\ (\bibinfo  {publisher} {Oxford University Press},\
  \bibinfo {address} {Oxford},\ \bibinfo {year} {2002})\BibitemShut {NoStop}%
\bibitem [{\citenamefont {Bu{\v{c}}a}\ and\ \citenamefont
  {Prosen}(2012)}]{BucaProsen2012}%
  \BibitemOpen
  \bibfield  {author} {\bibinfo {author} {\bibfnamefont {B.}~\bibnamefont
  {Bu{\v{c}}a}}\ and\ \bibinfo {author} {\bibfnamefont {T.}~\bibnamefont
  {Prosen}},\ }\emph {A note on symmetry reductions of the Lindblad equation:
  transport in constrained open spin chains},\ \href {\doibase
  10.1088/1367-2630/14/7/073007} {\bibfield  {journal} {\bibinfo  {journal}
  {New Journal of Physics}\ }\textbf {\bibinfo {volume} {14}},\ \bibinfo
  {pages} {073007} (\bibinfo {year} {2012})}\BibitemShut {NoStop}%
\bibitem{supp}
See Supplemental Material [url] regarding the details on the time-dependent matrix product states approach (tMPS), the model we consider in the case of a classical superlattice potential, the results for the case of a classical field under the action of a stochastic noise, the analytical derivations of the eigenvalues and eigenstates of the Liouvillian in the limit of vanishing hopping, the momentum distributions, the dynamics of single particle and pair correlations for an alternative initial state,  and the dynamics of quantum trajectories. The Supplemental Material includes Refs.~\cite{HalatiPhd,DalibardMolmer1992,GardinerZoller1992,Daley2014, WhiteFeiguin2004,DaleyVidal2004,Schollwoeck2011,StoudenmireWhite2010,WallRey2016, FishmanStoudenmire2020,NagyDomokos2008, SchuetzMorigi2013}
\bibitem [{\citenamefont {Albert}(2019)}]{Albert2019}%
  \BibitemOpen
  \bibfield  {author} {\bibinfo {author} {\bibfnamefont {V.~V.}\ \bibnamefont
  {Albert}},\ }\emph {Asymptotics of quantum channels: conserved quantities, an
  adiabatic limit, and matrix product states},\ \href {\doibase
  10.22331/q-2019-06-06-151} {\bibfield  {journal} {\bibinfo  {journal}
  {{Quantum}}\ }\textbf {\bibinfo {volume} {3}},\ \bibinfo {pages} {151}
  (\bibinfo {year} {2019})}\BibitemShut {NoStop}%
\bibitem [{\citenamefont {Halati}\ \emph {et~al.}(2022)\citenamefont {Halati},
  \citenamefont {Sheikhan},\ and\ \citenamefont {Kollath}}]{HalatiKollath2022}%
  \BibitemOpen
  \bibfield  {author} {\bibinfo {author} {\bibfnamefont {C.-M.}\ \bibnamefont
  {Halati}}, \bibinfo {author} {\bibfnamefont {A.}~\bibnamefont {Sheikhan}}, \
  and\ \bibinfo {author} {\bibfnamefont {C.}~\bibnamefont {Kollath}},\ }\emph
  {Breaking strong symmetries in dissipative quantum systems: Bosonic atoms
  coupled to a cavity},\ \href {\doibase 10.1103/PhysRevResearch.4.L012015}
  {\bibfield  {journal} {\bibinfo  {journal} {Phys. Rev. Research}\ }\textbf
  {\bibinfo {volume} {4}},\ \bibinfo {pages} {L012015} (\bibinfo {year}
  {2022})}\BibitemShut {NoStop}%
\bibitem [{\citenamefont {Maschler}\ \emph {et~al.}(2008)\citenamefont
  {Maschler}, \citenamefont {Mekhov},\ and\ \citenamefont
  {Ritsch}}]{MaschlerRitsch2008}%
  \BibitemOpen
  \bibfield  {author} {\bibinfo {author} {\bibfnamefont {C.}~\bibnamefont
  {Maschler}}, \bibinfo {author} {\bibfnamefont {I.~B.}\ \bibnamefont
  {Mekhov}}, \ and\ \bibinfo {author} {\bibfnamefont {H.}~\bibnamefont
  {Ritsch}},\ }\emph {Ultracold atoms in optical lattices generated by
  quantized light fields},\ \href {\doibase 10.1140/epjd/e2008-00016-4}
  {\bibfield  {journal} {\bibinfo  {journal} {The European Physical Journal D}\
  }\textbf {\bibinfo {volume} {46}},\ \bibinfo {pages} {545} (\bibinfo {year}
  {2008})}\BibitemShut {NoStop}%
\bibitem [{\citenamefont {Ritsch}\ \emph {et~al.}(2013)\citenamefont {Ritsch},
  \citenamefont {Domokos}, \citenamefont {Brennecke},\ and\ \citenamefont
  {Esslinger}}]{RitschEsslinger2013}%
  \BibitemOpen
  \bibfield  {author} {\bibinfo {author} {\bibfnamefont {H.}~\bibnamefont
  {Ritsch}}, \bibinfo {author} {\bibfnamefont {P.}~\bibnamefont {Domokos}},
  \bibinfo {author} {\bibfnamefont {F.}~\bibnamefont {Brennecke}}, \ and\
  \bibinfo {author} {\bibfnamefont {T.}~\bibnamefont {Esslinger}},\ }\emph
  {Cold atoms in cavity-generated dynamical optical potentials},\ \href
  {\doibase 10.1103/RevModPhys.85.553} {\bibfield  {journal} {\bibinfo
  {journal} {Rev. Mod. Phys.}\ }\textbf {\bibinfo {volume} {85}},\ \bibinfo
  {pages} {553} (\bibinfo {year} {2013})}\BibitemShut {NoStop}%
\bibitem [{\citenamefont {Mivehvar}\ \emph {et~al.}(2021)\citenamefont
  {Mivehvar}, \citenamefont {Piazza}, \citenamefont {Donner},\ and\
  \citenamefont {Ritsch}}]{MivehvarRitsch2021}%
  \BibitemOpen
  \bibfield  {author} {\bibinfo {author} {\bibfnamefont {F.}~\bibnamefont
  {Mivehvar}}, \bibinfo {author} {\bibfnamefont {F.}~\bibnamefont {Piazza}},
  \bibinfo {author} {\bibfnamefont {T.}~\bibnamefont {Donner}}, \ and\ \bibinfo
  {author} {\bibfnamefont {H.}~\bibnamefont {Ritsch}},\ }\emph {Cavity QED with
  quantum gases: new paradigms in many-body physics},\ \href {\doibase
  10.1080/00018732.2021.1969727} {\bibfield  {journal} {\bibinfo  {journal}
  {Advances in Physics}\ }\textbf {\bibinfo {volume} {70}},\ \bibinfo {pages}
  {1} (\bibinfo {year} {2021})}\BibitemShut {NoStop}%
\bibitem [{\citenamefont {Garc\'{\i}a-Ripoll}\ \emph
  {et~al.}(2009)\citenamefont {Garc\'{\i}a-Ripoll}, \citenamefont {D\"urr},
  \citenamefont {Syassen}, \citenamefont {Bauer}, \citenamefont {Lettner},
  \citenamefont {Rempe},\ and\ \citenamefont {Cirac}}]{Garcia-RipollCirac2009}%
  \BibitemOpen
  \bibfield  {author} {\bibinfo {author} {\bibfnamefont {J.~J.}\ \bibnamefont
  {Garc\'{\i}a-Ripoll}}, \bibinfo {author} {\bibfnamefont {S.}~\bibnamefont
  {D\"urr}}, \bibinfo {author} {\bibfnamefont {N.}~\bibnamefont {Syassen}},
  \bibinfo {author} {\bibfnamefont {D.~M.}\ \bibnamefont {Bauer}}, \bibinfo
  {author} {\bibfnamefont {M.}~\bibnamefont {Lettner}}, \bibinfo {author}
  {\bibfnamefont {G.}~\bibnamefont {Rempe}}, \ and\ \bibinfo {author}
  {\bibfnamefont {J.~I.}\ \bibnamefont {Cirac}},\ }\emph {Dissipation-induced
  hard-core boson gas in an optical lattice},\ \href
  {http://stacks.iop.org/1367-2630/11/i=1/a=013053} {\bibfield  {journal}
  {\bibinfo  {journal} {New Journal of Physics}\ }\textbf {\bibinfo {volume}
  {11}},\ \bibinfo {pages} {013053} (\bibinfo {year} {2009})}\BibitemShut
  {NoStop}%
\bibitem [{\citenamefont {Zanardi}\ and\ \citenamefont
  {Campos~Venuti}(2014)}]{ZanardiCamposVenuti2014}%
  \BibitemOpen
  \bibfield  {author} {\bibinfo {author} {\bibfnamefont {P.}~\bibnamefont
  {Zanardi}}\ and\ \bibinfo {author} {\bibfnamefont {L.}~\bibnamefont
  {Campos~Venuti}},\ }\emph {Coherent Quantum Dynamics in Steady-State
  Manifolds of Strongly Dissipative Systems},\ \href {\doibase
  10.1103/PhysRevLett.113.240406} {\bibfield  {journal} {\bibinfo  {journal}
  {Phys. Rev. Lett.}\ }\textbf {\bibinfo {volume} {113}},\ \bibinfo {pages}
  {240406} (\bibinfo {year} {2014})}\BibitemShut {NoStop}%
\bibitem [{\citenamefont {Jin}\ \emph {et~al.}(2024)\citenamefont {Jin},
  \citenamefont {Qiu},\ and\ \citenamefont {Ma}}]{JinMa2024}%
  \BibitemOpen
  \bibfield  {author} {\bibinfo {author} {\bibfnamefont {Y.-D.}\ \bibnamefont
  {Jin}}, \bibinfo {author} {\bibfnamefont {C.-D.}\ \bibnamefont {Qiu}}, \ and\
  \bibinfo {author} {\bibfnamefont {W.-L.}\ \bibnamefont {Ma}},\ }\emph {Theory
  of metastability in discrete-time open quantum dynamics},\ \href {\doibase
  10.1103/PhysRevA.109.042204} {\bibfield  {journal} {\bibinfo  {journal}
  {Phys. Rev. A}\ }\textbf {\bibinfo {volume} {109}},\ \bibinfo {pages}
  {042204} (\bibinfo {year} {2024})}\BibitemShut {NoStop}%
\bibitem [{\citenamefont {Cummings}(1965)}]{Cummings1965}%
  \BibitemOpen
  \bibfield  {author} {\bibinfo {author} {\bibfnamefont {F.~W.}\ \bibnamefont
  {Cummings}},\ }\emph {Stimulated Emission of Radiation in a Single Mode},\
  \href {\doibase 10.1103/PhysRev.140.A1051} {\bibfield  {journal} {\bibinfo
  {journal} {Phys. Rev.}\ }\textbf {\bibinfo {volume} {140}},\ \bibinfo {pages}
  {A1051} (\bibinfo {year} {1965})}\BibitemShut {NoStop}%
\bibitem [{\citenamefont {Eberly}\ \emph {et~al.}(1980)\citenamefont {Eberly},
  \citenamefont {Narozhny},\ and\ \citenamefont
  {Sanchez-Mondragon}}]{EberlySanchez-Mondragon1980}%
  \BibitemOpen
  \bibfield  {author} {\bibinfo {author} {\bibfnamefont {J.~H.}\ \bibnamefont
  {Eberly}}, \bibinfo {author} {\bibfnamefont {N.~B.}\ \bibnamefont
  {Narozhny}}, \ and\ \bibinfo {author} {\bibfnamefont {J.~J.}\ \bibnamefont
  {Sanchez-Mondragon}},\ }\emph {Periodic Spontaneous Collapse and Revival in a
  Simple Quantum Model},\ \href {\doibase 10.1103/PhysRevLett.44.1323}
  {\bibfield  {journal} {\bibinfo  {journal} {Phys. Rev. Lett.}\ }\textbf
  {\bibinfo {volume} {44}},\ \bibinfo {pages} {1323} (\bibinfo {year}
  {1980})}\BibitemShut {NoStop}%
\bibitem [{\citenamefont {Wright}\ \emph {et~al.}(1996)\citenamefont {Wright},
  \citenamefont {Walls},\ and\ \citenamefont {Garrison}}]{WrightGarrison1996}%
  \BibitemOpen
  \bibfield  {author} {\bibinfo {author} {\bibfnamefont {E.~M.}\ \bibnamefont
  {Wright}}, \bibinfo {author} {\bibfnamefont {D.~F.}\ \bibnamefont {Walls}}, \
  and\ \bibinfo {author} {\bibfnamefont {J.~C.}\ \bibnamefont {Garrison}},\
  }\emph {Collapses and Revivals of Bose-Einstein Condensates Formed in Small
  Atomic Samples},\ \href {\doibase 10.1103/PhysRevLett.77.2158} {\bibfield
  {journal} {\bibinfo  {journal} {Phys. Rev. Lett.}\ }\textbf {\bibinfo
  {volume} {77}},\ \bibinfo {pages} {2158} (\bibinfo {year}
  {1996})}\BibitemShut {NoStop}%
\bibitem [{\citenamefont {Dooley}\ \emph {et~al.}(2013)\citenamefont {Dooley},
  \citenamefont {McCrossan}, \citenamefont {Harland}, \citenamefont {Everitt},\
  and\ \citenamefont {Spiller}}]{DooleySpiller2013}%
  \BibitemOpen
  \bibfield  {author} {\bibinfo {author} {\bibfnamefont {S.}~\bibnamefont
  {Dooley}}, \bibinfo {author} {\bibfnamefont {F.}~\bibnamefont {McCrossan}},
  \bibinfo {author} {\bibfnamefont {D.}~\bibnamefont {Harland}}, \bibinfo
  {author} {\bibfnamefont {M.~J.}\ \bibnamefont {Everitt}}, \ and\ \bibinfo
  {author} {\bibfnamefont {T.~P.}\ \bibnamefont {Spiller}},\ }\emph {Collapse
  and revival and cat states with an $N$-spin system},\ \href {\doibase
  10.1103/PhysRevA.87.052323} {\bibfield  {journal} {\bibinfo  {journal} {Phys.
  Rev. A}\ }\textbf {\bibinfo {volume} {87}},\ \bibinfo {pages} {052323}
  (\bibinfo {year} {2013})}\BibitemShut {NoStop}%
\bibitem [{\citenamefont {Greiner}\ \emph {et~al.}(2002)\citenamefont
  {Greiner}, \citenamefont {Mandel}, \citenamefont {H{\"a}nsch},\ and\
  \citenamefont {Bloch}}]{GreinerBloch2002b}%
  \BibitemOpen
  \bibfield  {author} {\bibinfo {author} {\bibfnamefont {M.}~\bibnamefont
  {Greiner}}, \bibinfo {author} {\bibfnamefont {O.}~\bibnamefont {Mandel}},
  \bibinfo {author} {\bibfnamefont {T.~W.}\ \bibnamefont {H{\"a}nsch}}, \ and\
  \bibinfo {author} {\bibfnamefont {I.}~\bibnamefont {Bloch}},\ }\emph
  {Collapse and revival of the matter wave field of a Bose--Einstein
  condensate},\ \href {\doibase 10.1038/nature00968} {\bibfield  {journal}
  {\bibinfo  {journal} {Nature}\ }\textbf {\bibinfo {volume} {419}},\ \bibinfo
  {pages} {51} (\bibinfo {year} {2002})}\BibitemShut {NoStop}%
\bibitem [{\citenamefont {Goldblatt}\ \emph {et~al.}(2024)\citenamefont
  {Goldblatt}, \citenamefont {Martin},\ and\ \citenamefont
  {Wood}}]{GoldblattWood2024}%
  \BibitemOpen
  \bibfield  {author} {\bibinfo {author} {\bibfnamefont {R.}~\bibnamefont
  {Goldblatt}}, \bibinfo {author} {\bibfnamefont {A.}~\bibnamefont {Martin}}, \
  and\ \bibinfo {author} {\bibfnamefont {A.}~\bibnamefont {Wood}},\ }\emph
  {Sensing Coherent Nuclear Spin Dynamics with an Ensemble of Paramagnetic
  Nitrogen Spins},\ \href {\doibase 10.1103/PRXQuantum.5.020334} {\bibfield
  {journal} {\bibinfo  {journal} {PRX Quantum}\ }\textbf {\bibinfo {volume}
  {5}},\ \bibinfo {pages} {020334} (\bibinfo {year} {2024})}\BibitemShut
  {NoStop}%
\bibitem [{\citenamefont {Klinder}\ \emph {et~al.}(2015)\citenamefont
  {Klinder}, \citenamefont {Ke\ss{}ler}, \citenamefont {Bakhtiari},
  \citenamefont {Thorwart},\ and\ \citenamefont
  {Hemmerich}}]{KlinderHemmerich2015b}%
  \BibitemOpen
  \bibfield  {author} {\bibinfo {author} {\bibfnamefont {J.}~\bibnamefont
  {Klinder}}, \bibinfo {author} {\bibfnamefont {H.}~\bibnamefont {Ke\ss{}ler}},
  \bibinfo {author} {\bibfnamefont {M.~R.}\ \bibnamefont {Bakhtiari}}, \bibinfo
  {author} {\bibfnamefont {M.}~\bibnamefont {Thorwart}}, \ and\ \bibinfo
  {author} {\bibfnamefont {A.}~\bibnamefont {Hemmerich}},\ }\emph {Observation
  of a Superradiant Mott Insulator in the Dicke-Hubbard Model},\ \href
  {\doibase 10.1103/PhysRevLett.115.230403} {\bibfield  {journal} {\bibinfo
  {journal} {Phys. Rev. Lett.}\ }\textbf {\bibinfo {volume} {115}},\ \bibinfo
  {pages} {230403} (\bibinfo {year} {2015})}\BibitemShut {NoStop}%
\bibitem [{\citenamefont {Landig}\ \emph {et~al.}(2016)\citenamefont {Landig},
  \citenamefont {Hruby}, \citenamefont {Dogra}, \citenamefont {Landini},
  \citenamefont {Mottl}, \citenamefont {Donner},\ and\ \citenamefont
  {Esslinger}}]{LandigEsslinger2016}%
  \BibitemOpen
  \bibfield  {author} {\bibinfo {author} {\bibfnamefont {R.}~\bibnamefont
  {Landig}}, \bibinfo {author} {\bibfnamefont {L.}~\bibnamefont {Hruby}},
  \bibinfo {author} {\bibfnamefont {N.}~\bibnamefont {Dogra}}, \bibinfo
  {author} {\bibfnamefont {M.}~\bibnamefont {Landini}}, \bibinfo {author}
  {\bibfnamefont {R.}~\bibnamefont {Mottl}}, \bibinfo {author} {\bibfnamefont
  {T.}~\bibnamefont {Donner}}, \ and\ \bibinfo {author} {\bibfnamefont
  {T.}~\bibnamefont {Esslinger}},\ }\emph {{Quantum phases from competing
  short- and long-range interactions in an optical lattice}},\ \href
  {http://dx.doi.org/10.1038/nature17409} {\bibfield  {journal} {\bibinfo
  {journal} {Nature}\ }\textbf {\bibinfo {volume} {532}},\ \bibinfo {pages}
  {476} (\bibinfo {year} {2016})},\ \bibinfo {note} {letter}\BibitemShut
  {NoStop}%
\bibitem [{\citenamefont {Hruby}\ \emph {et~al.}(2018)\citenamefont {Hruby},
  \citenamefont {Dogra}, \citenamefont {Landini}, \citenamefont {Donner},\ and\
  \citenamefont {Esslinger}}]{HrubyEsslinger2018}%
  \BibitemOpen
  \bibfield  {author} {\bibinfo {author} {\bibfnamefont {L.}~\bibnamefont
  {Hruby}}, \bibinfo {author} {\bibfnamefont {N.}~\bibnamefont {Dogra}},
  \bibinfo {author} {\bibfnamefont {M.}~\bibnamefont {Landini}}, \bibinfo
  {author} {\bibfnamefont {T.}~\bibnamefont {Donner}}, \ and\ \bibinfo {author}
  {\bibfnamefont {T.}~\bibnamefont {Esslinger}},\ }\emph {Metastability and
  avalanche dynamics in strongly correlated gases with long-range
  interactions},\ \href {\doibase 10.1073/pnas.1720415115} {\bibfield
  {journal} {\bibinfo  {journal} {Proceedings of the National Academy of
  Sciences}\ }\textbf {\bibinfo {volume} {115}},\ \bibinfo {pages} {3279}
  (\bibinfo {year} {2018})}\BibitemShut {NoStop}%
\bibitem [{\citenamefont {Niedenzu}\ \emph {et~al.}(2010)\citenamefont
  {Niedenzu}, \citenamefont {Schulze}, \citenamefont {Vukics},\ and\
  \citenamefont {Ritsch}}]{NiedenzuRitsch2010}%
  \BibitemOpen
  \bibfield  {author} {\bibinfo {author} {\bibfnamefont {W.}~\bibnamefont
  {Niedenzu}}, \bibinfo {author} {\bibfnamefont {R.}~\bibnamefont {Schulze}},
  \bibinfo {author} {\bibfnamefont {A.}~\bibnamefont {Vukics}}, \ and\ \bibinfo
  {author} {\bibfnamefont {H.}~\bibnamefont {Ritsch}},\ }\emph {Microscopic
  dynamics of ultracold particles in a ring-cavity optical lattice},\ \href
  {\doibase 10.1103/PhysRevA.82.043605} {\bibfield  {journal} {\bibinfo
  {journal} {Phys. Rev. A}\ }\textbf {\bibinfo {volume} {82}},\ \bibinfo
  {pages} {043605} (\bibinfo {year} {2010})}\BibitemShut {NoStop}%
\bibitem [{\citenamefont {Silver}\ \emph {et~al.}(2010)\citenamefont {Silver},
  \citenamefont {Hohenadler}, \citenamefont {Bhaseen},\ and\ \citenamefont
  {Simons}}]{SilverSimons2010}%
  \BibitemOpen
  \bibfield  {author} {\bibinfo {author} {\bibfnamefont {A.~O.}\ \bibnamefont
  {Silver}}, \bibinfo {author} {\bibfnamefont {M.}~\bibnamefont {Hohenadler}},
  \bibinfo {author} {\bibfnamefont {M.~J.}\ \bibnamefont {Bhaseen}}, \ and\
  \bibinfo {author} {\bibfnamefont {B.~D.}\ \bibnamefont {Simons}},\ }\emph
  {Bose-Hubbard models coupled to cavity light fields},\ \href {\doibase
  10.1103/PhysRevA.81.023617} {\bibfield  {journal} {\bibinfo  {journal} {Phys.
  Rev. A}\ }\textbf {\bibinfo {volume} {81}},\ \bibinfo {pages} {023617}
  (\bibinfo {year} {2010})}\BibitemShut {NoStop}%
\bibitem [{\citenamefont {Fern\'andez-Vidal}\ \emph {et~al.}(2010)\citenamefont
  {Fern\'andez-Vidal}, \citenamefont {De~Chiara}, \citenamefont {Larson},\ and\
  \citenamefont {Morigi}}]{VidalMorigi2010}%
  \BibitemOpen
  \bibfield  {author} {\bibinfo {author} {\bibfnamefont {S.}~\bibnamefont
  {Fern\'andez-Vidal}}, \bibinfo {author} {\bibfnamefont {G.}~\bibnamefont
  {De~Chiara}}, \bibinfo {author} {\bibfnamefont {J.}~\bibnamefont {Larson}}, \
  and\ \bibinfo {author} {\bibfnamefont {G.}~\bibnamefont {Morigi}},\ }\emph
  {Quantum ground state of self-organized atomic crystals in optical
  resonators},\ \href {\doibase 10.1103/PhysRevA.81.043407} {\bibfield
  {journal} {\bibinfo  {journal} {Phys. Rev. A}\ }\textbf {\bibinfo {volume}
  {81}},\ \bibinfo {pages} {043407} (\bibinfo {year} {2010})}\BibitemShut
  {NoStop}%
\bibitem [{\citenamefont {Li}\ \emph {et~al.}(2013)\citenamefont {Li},
  \citenamefont {He},\ and\ \citenamefont {Hofstetter}}]{LiHofstetter2013}%
  \BibitemOpen
  \bibfield  {author} {\bibinfo {author} {\bibfnamefont {Y.}~\bibnamefont
  {Li}}, \bibinfo {author} {\bibfnamefont {L.}~\bibnamefont {He}}, \ and\
  \bibinfo {author} {\bibfnamefont {W.}~\bibnamefont {Hofstetter}},\ }\emph
  {Lattice-supersolid phase of strongly correlated bosons in an optical
  cavity},\ \href {\doibase 10.1103/PhysRevA.87.051604} {\bibfield  {journal}
  {\bibinfo  {journal} {Phys. Rev. A}\ }\textbf {\bibinfo {volume} {87}},\
  \bibinfo {pages} {051604} (\bibinfo {year} {2013})}\BibitemShut {NoStop}%
\bibitem [{\citenamefont {Bakhtiari}\ \emph {et~al.}(2015)\citenamefont
  {Bakhtiari}, \citenamefont {Hemmerich}, \citenamefont {Ritsch},\ and\
  \citenamefont {Thorwart}}]{BakhtiariThorwart2015}%
  \BibitemOpen
  \bibfield  {author} {\bibinfo {author} {\bibfnamefont {M.~R.}\ \bibnamefont
  {Bakhtiari}}, \bibinfo {author} {\bibfnamefont {A.}~\bibnamefont
  {Hemmerich}}, \bibinfo {author} {\bibfnamefont {H.}~\bibnamefont {Ritsch}}, \
  and\ \bibinfo {author} {\bibfnamefont {M.}~\bibnamefont {Thorwart}},\ }\emph
  {Nonequilibrium Phase Transition of Interacting Bosons in an Intra-Cavity
  Optical Lattice},\ \href {\doibase 10.1103/PhysRevLett.114.123601} {\bibfield
   {journal} {\bibinfo  {journal} {Phys. Rev. Lett.}\ }\textbf {\bibinfo
  {volume} {114}},\ \bibinfo {pages} {123601} (\bibinfo {year}
  {2015})}\BibitemShut {NoStop}%
\bibitem [{\citenamefont {Flottat}\ \emph {et~al.}(2017)\citenamefont
  {Flottat}, \citenamefont {de~Parny}, \citenamefont {H\'ebert}, \citenamefont
  {Rousseau},\ and\ \citenamefont {Batrouni}}]{FlottatBatrouni2017}%
  \BibitemOpen
  \bibfield  {author} {\bibinfo {author} {\bibfnamefont {T.}~\bibnamefont
  {Flottat}}, \bibinfo {author} {\bibfnamefont {L.~d.~F.}\ \bibnamefont
  {de~Parny}}, \bibinfo {author} {\bibfnamefont {F.}~\bibnamefont {H\'ebert}},
  \bibinfo {author} {\bibfnamefont {V.~G.}\ \bibnamefont {Rousseau}}, \ and\
  \bibinfo {author} {\bibfnamefont {G.~G.}\ \bibnamefont {Batrouni}},\ }\emph
  {Phase diagram of bosons in a two-dimensional optical lattice with
  infinite-range cavity-mediated interactions},\ \href {\doibase
  10.1103/PhysRevB.95.144501} {\bibfield  {journal} {\bibinfo  {journal} {Phys.
  Rev. B}\ }\textbf {\bibinfo {volume} {95}},\ \bibinfo {pages} {144501}
  (\bibinfo {year} {2017})}\BibitemShut {NoStop}%
\bibitem [{\citenamefont {Rodr\'{\i}guez~Chiacchio}\ and\ \citenamefont
  {Nunnenkamp}(2018)}]{ChiacchioNunnenkamp2018}%
  \BibitemOpen
  \bibfield  {author} {\bibinfo {author} {\bibfnamefont {E.~I.}\ \bibnamefont
  {Rodr\'{\i}guez~Chiacchio}}\ and\ \bibinfo {author} {\bibfnamefont
  {A.}~\bibnamefont {Nunnenkamp}},\ }\emph {Tuning the relaxation dynamics of
  ultracold atoms in a lattice with an optical cavity},\ \href {\doibase
  10.1103/PhysRevA.97.033618} {\bibfield  {journal} {\bibinfo  {journal} {Phys.
  Rev. A}\ }\textbf {\bibinfo {volume} {97}},\ \bibinfo {pages} {033618}
  (\bibinfo {year} {2018})}\BibitemShut {NoStop}%
\bibitem [{\citenamefont {Lin}\ \emph {et~al.}(2019)\citenamefont {Lin},
  \citenamefont {Papariello}, \citenamefont {Molignini}, \citenamefont
  {Chitra},\ and\ \citenamefont {Lode}}]{LinLode2019}%
  \BibitemOpen
  \bibfield  {author} {\bibinfo {author} {\bibfnamefont {R.}~\bibnamefont
  {Lin}}, \bibinfo {author} {\bibfnamefont {L.}~\bibnamefont {Papariello}},
  \bibinfo {author} {\bibfnamefont {P.}~\bibnamefont {Molignini}}, \bibinfo
  {author} {\bibfnamefont {R.}~\bibnamefont {Chitra}}, \ and\ \bibinfo {author}
  {\bibfnamefont {A.~U.~J.}\ \bibnamefont {Lode}},\ }\emph
  {Superfluid--Mott-insulator transition of ultracold superradiant bosons in a
  cavity},\ \href {\doibase 10.1103/PhysRevA.100.013611} {\bibfield  {journal}
  {\bibinfo  {journal} {Phys. Rev. A}\ }\textbf {\bibinfo {volume} {100}},\
  \bibinfo {pages} {013611} (\bibinfo {year} {2019})}\BibitemShut {NoStop}%
\bibitem [{\citenamefont {Himbert}\ \emph {et~al.}(2019)\citenamefont
  {Himbert}, \citenamefont {Cormick}, \citenamefont {Kraus}, \citenamefont
  {Sharma},\ and\ \citenamefont {Morigi}}]{HimbertMorigi2019}%
  \BibitemOpen
  \bibfield  {author} {\bibinfo {author} {\bibfnamefont {L.}~\bibnamefont
  {Himbert}}, \bibinfo {author} {\bibfnamefont {C.}~\bibnamefont {Cormick}},
  \bibinfo {author} {\bibfnamefont {R.}~\bibnamefont {Kraus}}, \bibinfo
  {author} {\bibfnamefont {S.}~\bibnamefont {Sharma}}, \ and\ \bibinfo {author}
  {\bibfnamefont {G.}~\bibnamefont {Morigi}},\ }\emph {Mean-field phase diagram
  of the extended Bose-Hubbard model of many-body cavity quantum
  electrodynamics},\ \href {\doibase 10.1103/PhysRevA.99.043633} {\bibfield
  {journal} {\bibinfo  {journal} {Phys. Rev. A}\ }\textbf {\bibinfo {volume}
  {99}},\ \bibinfo {pages} {043633} (\bibinfo {year} {2019})}\BibitemShut
  {NoStop}%
\bibitem [{\citenamefont {Halati}\ \emph
  {et~al.}(2020{\natexlab{a}})\citenamefont {Halati}, \citenamefont {Sheikhan},
  \citenamefont {Ritsch},\ and\ \citenamefont {Kollath}}]{HalatiKollath2020}%
  \BibitemOpen
  \bibfield  {author} {\bibinfo {author} {\bibfnamefont {C.-M.}\ \bibnamefont
  {Halati}}, \bibinfo {author} {\bibfnamefont {A.}~\bibnamefont {Sheikhan}},
  \bibinfo {author} {\bibfnamefont {H.}~\bibnamefont {Ritsch}}, \ and\ \bibinfo
  {author} {\bibfnamefont {C.}~\bibnamefont {Kollath}},\ }\emph {Numerically
  Exact Treatment of Many-Body Self-Organization in a Cavity},\ \href {\doibase
  10.1103/PhysRevLett.125.093604} {\bibfield  {journal} {\bibinfo  {journal}
  {Phys. Rev. Lett.}\ }\textbf {\bibinfo {volume} {125}},\ \bibinfo {pages}
  {093604} (\bibinfo {year} {2020}{\natexlab{a}})}\BibitemShut {NoStop}%
\bibitem [{\citenamefont {Bezvershenko}\ \emph {et~al.}(2021)\citenamefont
  {Bezvershenko}, \citenamefont {Halati}, \citenamefont {Sheikhan},
  \citenamefont {Kollath},\ and\ \citenamefont
  {Rosch}}]{BezvershenkoRosch2021}%
  \BibitemOpen
  \bibfield  {author} {\bibinfo {author} {\bibfnamefont {A.~V.}\ \bibnamefont
  {Bezvershenko}}, \bibinfo {author} {\bibfnamefont {C.-M.}\ \bibnamefont
  {Halati}}, \bibinfo {author} {\bibfnamefont {A.}~\bibnamefont {Sheikhan}},
  \bibinfo {author} {\bibfnamefont {C.}~\bibnamefont {Kollath}}, \ and\
  \bibinfo {author} {\bibfnamefont {A.}~\bibnamefont {Rosch}},\ }\emph {Dicke
  Transition in Open Many-Body Systems Determined by Fluctuation Effects},\
  \href {\doibase 10.1103/PhysRevLett.127.173606} {\bibfield  {journal}
  {\bibinfo  {journal} {Phys. Rev. Lett.}\ }\textbf {\bibinfo {volume} {127}},\
  \bibinfo {pages} {173606} (\bibinfo {year} {2021})}\BibitemShut {NoStop}%
\bibitem [{\citenamefont {Sharma}\ \emph {et~al.}(2022)\citenamefont {Sharma},
  \citenamefont {J\"ager}, \citenamefont {Kraus}, \citenamefont {Roscilde},\
  and\ \citenamefont {Morigi}}]{SharmaMorigi2022}%
  \BibitemOpen
  \bibfield  {author} {\bibinfo {author} {\bibfnamefont {S.}~\bibnamefont
  {Sharma}}, \bibinfo {author} {\bibfnamefont {S.~B.}\ \bibnamefont {J\"ager}},
  \bibinfo {author} {\bibfnamefont {R.}~\bibnamefont {Kraus}}, \bibinfo
  {author} {\bibfnamefont {T.}~\bibnamefont {Roscilde}}, \ and\ \bibinfo
  {author} {\bibfnamefont {G.}~\bibnamefont {Morigi}},\ }\emph {Quantum
  Critical Behavior of Entanglement in Lattice Bosons with Cavity-Mediated
  Long-Range Interactions},\ \href {\doibase 10.1103/PhysRevLett.129.143001}
  {\bibfield  {journal} {\bibinfo  {journal} {Phys. Rev. Lett.}\ }\textbf
  {\bibinfo {volume} {129}},\ \bibinfo {pages} {143001} (\bibinfo {year}
  {2022})}\BibitemShut {NoStop}%
\bibitem [{\citenamefont {Chanda}\ \emph {et~al.}(2022)\citenamefont {Chanda},
  \citenamefont {Kraus}, \citenamefont {Zakrzewski},\ and\ \citenamefont
  {Morigi}}]{ChandaMorigi2022}%
  \BibitemOpen
  \bibfield  {author} {\bibinfo {author} {\bibfnamefont {T.}~\bibnamefont
  {Chanda}}, \bibinfo {author} {\bibfnamefont {R.}~\bibnamefont {Kraus}},
  \bibinfo {author} {\bibfnamefont {J.}~\bibnamefont {Zakrzewski}}, \ and\
  \bibinfo {author} {\bibfnamefont {G.}~\bibnamefont {Morigi}},\ }\emph {Bond
  order via cavity-mediated interactions},\ \href {\doibase
  10.1103/PhysRevB.106.075137} {\bibfield  {journal} {\bibinfo  {journal}
  {Phys. Rev. B}\ }\textbf {\bibinfo {volume} {106}},\ \bibinfo {pages}
  {075137} (\bibinfo {year} {2022})}\BibitemShut {NoStop}%
\bibitem [{\citenamefont {Halati}\ \emph
  {et~al.}(2020{\natexlab{b}})\citenamefont {Halati}, \citenamefont
  {Sheikhan},\ and\ \citenamefont {Kollath}}]{HalatiKollath2020b}%
  \BibitemOpen
  \bibfield  {author} {\bibinfo {author} {\bibfnamefont {C.-M.}\ \bibnamefont
  {Halati}}, \bibinfo {author} {\bibfnamefont {A.}~\bibnamefont {Sheikhan}}, \
  and\ \bibinfo {author} {\bibfnamefont {C.}~\bibnamefont {Kollath}},\ }\emph
  {Theoretical methods to treat a single dissipative bosonic mode coupled
  globally to an interacting many-body system},\ \href {\doibase
  10.1103/PhysRevResearch.2.043255} {\bibfield  {journal} {\bibinfo  {journal}
  {Phys. Rev. Research}\ }\textbf {\bibinfo {volume} {2}},\ \bibinfo {pages}
  {043255} (\bibinfo {year} {2020}{\natexlab{b}})}\BibitemShut {NoStop}%
\bibitem [{\citenamefont {S\'anchez Mu\~noz}\ \emph {et~al.}(2019)\citenamefont
  {S\'anchez Mu\~noz}, \citenamefont {Bu\ifmmode~\check{c}\else \v{c}\fi{}a},
  \citenamefont {Tindall}, \citenamefont {Gonz\'alez-Tudela}, \citenamefont
  {Jaksch},\ and\ \citenamefont {Porras}}]{MunozPorras2019}%
  \BibitemOpen
  \bibfield  {author} {\bibinfo {author} {\bibfnamefont {C.}~\bibnamefont
  {S\'anchez Mu\~noz}}, \bibinfo {author} {\bibfnamefont {B.}~\bibnamefont
  {Bu\ifmmode~\check{c}\else \v{c}\fi{}a}}, \bibinfo {author} {\bibfnamefont
  {J.}~\bibnamefont {Tindall}}, \bibinfo {author} {\bibfnamefont
  {A.}~\bibnamefont {Gonz\'alez-Tudela}}, \bibinfo {author} {\bibfnamefont
  {D.}~\bibnamefont {Jaksch}}, \ and\ \bibinfo {author} {\bibfnamefont
  {D.}~\bibnamefont {Porras}},\ }\emph {Symmetries and conservation laws in
  quantum trajectories: Dissipative freezing},\ \href {\doibase
  10.1103/PhysRevA.100.042113} {\bibfield  {journal} {\bibinfo  {journal}
  {Phys. Rev. A}\ }\textbf {\bibinfo {volume} {100}},\ \bibinfo {pages}
  {042113} (\bibinfo {year} {2019})}\BibitemShut {NoStop}%
\bibitem [{\citenamefont {Tindall}\ \emph {et~al.}(2023)\citenamefont
  {Tindall}, \citenamefont {Jaksch},\ and\ \citenamefont
  {Muñoz}}]{TindallMunoz2023}%
  \BibitemOpen
  \bibfield  {author} {\bibinfo {author} {\bibfnamefont {J.}~\bibnamefont
  {Tindall}}, \bibinfo {author} {\bibfnamefont {D.}~\bibnamefont {Jaksch}}, \
  and\ \bibinfo {author} {\bibfnamefont {C.~S.}\ \bibnamefont {Muñoz}},\
  }\emph {{On the generality of symmetry breaking and dissipative freezing in
  quantum trajectories}},\ \href {\doibase 10.21468/SciPostPhysCore.6.1.004}
  {\bibfield  {journal} {\bibinfo  {journal} {SciPost Phys. Core}\ }\textbf
  {\bibinfo {volume} {6}},\ \bibinfo {pages} {004} (\bibinfo {year}
  {2023})}\BibitemShut {NoStop}%
\bibitem{datazenodo}
C.-M.~Halati, A.~Sheikhan, G.~Morigi, and C.~Kollath, (2024). Figure data for article "Controlling the dynamics of atomic correlations via the coupling to a dissipative cavity" [Data set]. Zenodo. \href {https://doi.org/10.5281/zenodo.10890066} {10.5281/zenodo.10890066}
\bibitem [{\citenamefont {Halati}(2021)}]{HalatiPhd}%
  \BibitemOpen
  \bibfield  {author} {\bibinfo {author} {\bibfnamefont {C.-M.}\ \bibnamefont
  {Halati}},\ }\emph {\bibinfo {title} {External Control of Many-Body Quantum
  Systems}},\ \href
  {https://bonndoc.ulb.uni-bonn.de/xmlui/handle/20.500.11811/9343} {Ph.D.
  thesis},\ \bibinfo  {school} {University of Bonn} (\bibinfo {year}
  {2021})\BibitemShut {NoStop}%
\bibitem [{\citenamefont {Dalibard}\ \emph {et~al.}(1992)\citenamefont
  {Dalibard}, \citenamefont {Castin},\ and\ \citenamefont
  {M\o{}lmer}}]{DalibardMolmer1992}%
  \BibitemOpen
  \bibfield  {author} {\bibinfo {author} {\bibfnamefont {J.}~\bibnamefont
  {Dalibard}}, \bibinfo {author} {\bibfnamefont {Y.}~\bibnamefont {Castin}}, \
  and\ \bibinfo {author} {\bibfnamefont {K.}~\bibnamefont {M\o{}lmer}},\ }\emph
  {Wave-function approach to dissipative processes in quantum optics},\ \href
  {\doibase 10.1103/PhysRevLett.68.580} {\bibfield  {journal} {\bibinfo
  {journal} {Phys. Rev. Lett.}\ }\textbf {\bibinfo {volume} {68}},\ \bibinfo
  {pages} {580} (\bibinfo {year} {1992})}\BibitemShut {NoStop}%
\bibitem [{\citenamefont {Gardiner}\ \emph {et~al.}(1992)\citenamefont
  {Gardiner}, \citenamefont {Parkins},\ and\ \citenamefont
  {Zoller}}]{GardinerZoller1992}%
  \BibitemOpen
  \bibfield  {author} {\bibinfo {author} {\bibfnamefont {C.~W.}\ \bibnamefont
  {Gardiner}}, \bibinfo {author} {\bibfnamefont {A.~S.}\ \bibnamefont
  {Parkins}}, \ and\ \bibinfo {author} {\bibfnamefont {P.}~\bibnamefont
  {Zoller}},\ }\emph {Wave-function quantum stochastic differential equations
  and quantum-jump simulation methods},\ \href {\doibase
  10.1103/PhysRevA.46.4363} {\bibfield  {journal} {\bibinfo  {journal} {Phys.
  Rev. A}\ }\textbf {\bibinfo {volume} {46}},\ \bibinfo {pages} {4363}
  (\bibinfo {year} {1992})}\BibitemShut {NoStop}%
\bibitem [{\citenamefont {Daley}(2014)}]{Daley2014}%
  \BibitemOpen
  \bibfield  {author} {\bibinfo {author} {\bibfnamefont {A.~J.}\ \bibnamefont
  {Daley}},\ }\emph {Quantum trajectories and open many-body quantum systems},\
  \href {\doibase 10.1080/00018732.2014.933502} {\bibfield  {journal} {\bibinfo
   {journal} {Advances in Physics}\ }\textbf {\bibinfo {volume} {63}},\
  \bibinfo {pages} {77} (\bibinfo {year} {2014})}\BibitemShut {NoStop}%
\bibitem [{\citenamefont {White}\ and\ \citenamefont
  {Feiguin}(2004)}]{WhiteFeiguin2004}%
  \BibitemOpen
  \bibfield  {author} {\bibinfo {author} {\bibfnamefont {S.~R.}\ \bibnamefont
  {White}}\ and\ \bibinfo {author} {\bibfnamefont {A.~E.}\ \bibnamefont
  {Feiguin}},\ }\emph {Real-Time Evolution Using the Density Matrix
  Renormalization Group},\ \href {\doibase 10.1103/PhysRevLett.93.076401}
  {\bibfield  {journal} {\bibinfo  {journal} {Phys. Rev. Lett.}\ }\textbf
  {\bibinfo {volume} {93}},\ \bibinfo {pages} {076401} (\bibinfo {year}
  {2004})}\BibitemShut {NoStop}%
\bibitem [{\citenamefont {Daley}\ \emph {et~al.}(2004)\citenamefont {Daley},
  \citenamefont {Kollath}, \citenamefont {Schollwöck},\ and\ \citenamefont
  {Vidal}}]{DaleyVidal2004}%
  \BibitemOpen
  \bibfield  {author} {\bibinfo {author} {\bibfnamefont {A.~J.}\ \bibnamefont
  {Daley}}, \bibinfo {author} {\bibfnamefont {C.}~\bibnamefont {Kollath}},
  \bibinfo {author} {\bibfnamefont {U.}~\bibnamefont {Schollwöck}}, \ and\
  \bibinfo {author} {\bibfnamefont {G.}~\bibnamefont {Vidal}},\ }\emph
  {Time-dependent density-matrix renormalization-group using adaptive effective
  Hilbert spaces},\ \href {\doibase 10.1088/1742-5468/2004/04/P04005}
  {\bibfield  {journal} {\bibinfo  {journal} {Journal of Statistical Mechanics:
  Theory and Experiment}\ }\textbf {\bibinfo {volume} {2004}},\ \bibinfo
  {pages} {P04005} (\bibinfo {year} {2004})}\BibitemShut {NoStop}%
\bibitem [{\citenamefont {Schollw{\"o}ck}(2011)}]{Schollwoeck2011}%
  \BibitemOpen
  \bibfield  {author} {\bibinfo {author} {\bibfnamefont {U.}~\bibnamefont
  {Schollw{\"o}ck}},\ }\emph {The density-matrix renormalization group in the
  age of matrix product states},\ \href {\doibase
  http://dx.doi.org/10.1016/j.aop.2010.09.012} {\bibfield  {journal} {\bibinfo
  {journal} {Annals of Physics}\ }\textbf {\bibinfo {volume} {326}},\ \bibinfo
  {pages} {96 } (\bibinfo {year} {2011})}\BibitemShut {NoStop}%
\bibitem [{\citenamefont {Stoudenmire}\ and\ \citenamefont
  {White}(2010)}]{StoudenmireWhite2010}%
  \BibitemOpen
  \bibfield  {author} {\bibinfo {author} {\bibfnamefont {E.~M.}\ \bibnamefont
  {Stoudenmire}}\ and\ \bibinfo {author} {\bibfnamefont {S.~R.}\ \bibnamefont
  {White}},\ }\emph {Minimally entangled typical thermal state algorithms},\
  \href {\doibase 10.1088/1367-2630/12/5/055026} {\bibfield  {journal}
  {\bibinfo  {journal} {New Journal of Physics}\ }\textbf {\bibinfo {volume}
  {12}},\ \bibinfo {pages} {055026} (\bibinfo {year} {2010})}\BibitemShut
  {NoStop}%
\bibitem [{\citenamefont {Wall}\ \emph {et~al.}(2016)\citenamefont {Wall},
  \citenamefont {Safavi-Naini},\ and\ \citenamefont {Rey}}]{WallRey2016}%
  \BibitemOpen
  \bibfield  {author} {\bibinfo {author} {\bibfnamefont {M.~L.}\ \bibnamefont
  {Wall}}, \bibinfo {author} {\bibfnamefont {A.}~\bibnamefont {Safavi-Naini}},
  \ and\ \bibinfo {author} {\bibfnamefont {A.~M.}\ \bibnamefont {Rey}},\ }\emph
  {Simulating generic spin-boson models with matrix product states},\ \href
  {\doibase 10.1103/PhysRevA.94.053637} {\bibfield  {journal} {\bibinfo
  {journal} {Phys. Rev. A}\ }\textbf {\bibinfo {volume} {94}},\ \bibinfo
  {pages} {053637} (\bibinfo {year} {2016})}\BibitemShut {NoStop}%
\bibitem [{\citenamefont {Fishman}\ \emph {et~al.}(2022)\citenamefont
  {Fishman}, \citenamefont {White},\ and\ \citenamefont
  {Stoudenmire}}]{FishmanStoudenmire2020}%
  \BibitemOpen
  \bibfield  {author} {\bibinfo {author} {\bibfnamefont {M.}~\bibnamefont
  {Fishman}}, \bibinfo {author} {\bibfnamefont {S.~R.}\ \bibnamefont {White}},
  \ and\ \bibinfo {author} {\bibfnamefont {E.~M.}\ \bibnamefont
  {Stoudenmire}},\ }\emph {{The ITensor Software Library for Tensor Network
  Calculations}},\ \href {\doibase 10.21468/SciPostPhysCodeb.4} {\bibfield
  {journal} {\bibinfo  {journal} {SciPost Phys. Codebases}\ ,\ \bibinfo {pages}
  {4}} (\bibinfo {year} {2022})}\BibitemShut {NoStop}%
\bibitem [{\citenamefont {Nagy}\ \emph {et~al.}(2008)\citenamefont {Nagy},
  \citenamefont {Szirmai},\ and\ \citenamefont {Domokos}}]{NagyDomokos2008}%
  \BibitemOpen
  \bibfield  {author} {\bibinfo {author} {\bibfnamefont {D.}~\bibnamefont
  {Nagy}}, \bibinfo {author} {\bibfnamefont {G.}~\bibnamefont {Szirmai}}, \
  and\ \bibinfo {author} {\bibfnamefont {P.}~\bibnamefont {Domokos}},\ }\emph
  {Self-organization of a Bose-Einstein condensate in an optical cavity},\
  \href {\doibase 10.1140/epjd/e2008-00074-6} {\bibfield  {journal} {\bibinfo
  {journal} {The European Physical Journal D}\ }\textbf {\bibinfo {volume}
  {48}},\ \bibinfo {pages} {127} (\bibinfo {year} {2008})}\BibitemShut
  {NoStop}%
\bibitem [{\citenamefont {Sch\"utz}\ \emph {et~al.}(2013)\citenamefont
  {Sch\"utz}, \citenamefont {Habibian},\ and\ \citenamefont
  {Morigi}}]{SchuetzMorigi2013}%
  \BibitemOpen
  \bibfield  {author} {\bibinfo {author} {\bibfnamefont {S.}~\bibnamefont
  {Sch\"utz}}, \bibinfo {author} {\bibfnamefont {H.}~\bibnamefont {Habibian}},
  \ and\ \bibinfo {author} {\bibfnamefont {G.}~\bibnamefont {Morigi}},\ }\emph
  {Cooling of atomic ensembles in optical cavities: Semiclassical limit},\
  \href {\doibase 10.1103/PhysRevA.88.033427} {\bibfield  {journal} {\bibinfo
  {journal} {Phys. Rev. A}\ }\textbf {\bibinfo {volume} {88}},\ \bibinfo
  {pages} {033427} (\bibinfo {year} {2013})}\BibitemShut {NoStop}%
\end{thebibliography}
\end{document}